\documentclass[%
 reprint,
 superscriptaddress,
 amsmath,amssymb,
 aps,
 prl,
]{revtex4-1}

\usepackage{graphicx}
\usepackage{dcolumn}
\usepackage{bm}
\usepackage{colortbl}
\usepackage[table]{xcolor}
\usepackage{makecell}
\usepackage[mathscr]{euscript}
\usepackage[table]{xcolor}
\usepackage{multirow}
\usepackage{braket}
\usepackage[pagewise]{lineno}

\begin{document}


\title{Zeeman effect in centrosymmetric antiferromagnetic semiconductors controlled by electric field}

\author{Hong Jian Zhao}
 \affiliation{International Center for Computational Method and Software, College of Physics, Jilin University, Changchun 130012, China}
 \affiliation{Key Laboratory of Physics and Technology for Advanced Batteries (Ministry of Education), College of Physics, Jilin University, Changchun 130012, China}
 \affiliation{International Center of Future Science, Jilin University, Changchun 130012, China}
 \author{Xinran Liu}
 \affiliation{International Center for Computational Method and Software, College of Physics, Jilin University, Changchun 130012, China}
 \author{Yanchao Wang}
 \email{wyc@calypso.cn}
 \affiliation{International Center for Computational Method and Software, College of Physics, Jilin University, Changchun 130012, China}
 \affiliation{State Key Laboratory of Superhard Materials, College of Physics, Jilin University, Changchun 130012, China}
 \author{Yurong Yang}
\affiliation{National Laboratory of Solid State Microstructures and Jiangsu Key Laboratory of Artificial Functional Materials, Department of Materials Science and Engineering, Nanjing University, Nanjing 210093, China}
\author{Laurent Bellaiche}
\affiliation{Physics Department and Institute for Nanoscience and Engineering, University of Arkansas, Fayetteville, Arkansas 72701, USA}
\author{Yanming Ma}
\email{mym@jlu.edu.cn}
 \affiliation{International Center for Computational Method and Software, College of Physics, Jilin University, Changchun 130012, China}
 \affiliation{International Center of Future Science, Jilin University, Changchun 130012, China}
 \affiliation{State Key Laboratory of Superhard Materials, College of Physics, Jilin University, Changchun 130012, China}

\begin{abstract}
Centrosymmetric antiferromagnetic semiconductors, although abundant in nature, seem less promising than ferromagnets and ferroelectrics for practical applications in semiconductor spintronics. As a matter of fact, the lack of spontaneous polarization  and magnetization hinders the efficient utilization of electronic spin in these materials. 
Here, we propose a paradigm to harness electronic spin in centrosymmetric antiferromagnets via Zeeman spin splittings of electronic energy levels -- termed as spin Zeeman effect -- which is controlled by electric field.
By symmetry analysis, we identify twenty-one centrosymmetric antiferromagnetic point groups that accommodate such a spin Zeeman effect. We further predict by first-principles that two antiferromagnetic semiconductors, Fe$_2$TeO$_6$ and SrFe$_2$S$_2$O, are excellent candidates showcasing Zeeman splittings as large as $\sim$55 and $\sim$30 meV, respectively, induced by an electric field of 6 MV/cm.  Moreover, the electronic spin magnetization associated to the splitting energy levels can be switched by reversing the electric field. 
Our work thus sheds light on the electric-field control of electronic spin in antiferromagnets, which broadens the scope of application of centrosymmetric antiferromagnetic semiconductors.
\end{abstract}

\maketitle

\noindent
\textit{Introduction. --} In semiconductors, the creation of magnetically or electrically controllable spin splittings with relatively large magnitudes is at the heart of designing semiconductor spintronic devices (e.g., spin transistor)~\cite{spintronic,spintronic2,spinhall,rashbadresselhaus1,rashbadresselhaus2}. The conventional ferromagnetic or ferroelectric semiconductors naturally host such controllable spin splittings because of the existence of a spontaneous magnetization or polarization, thanks to Rashba-Dresselhaus~\cite{rashbadresselhaus1,rashbadresselhaus2,rashbajetp,dresselhausoriginal}, or Zeeman~\cite{zeemanmagnet,spintronic} effect. In sharp contrast, the centrosymmetric antiferromagnetic semiconductors do not have any polarization and their magnetization is either null or tiny. The lack of polarization and magnetization makes it challenging to generate a sizable and controllable spin splitting, by magnetic or electric field, in centrosymmetric antiferromagnetic semiconductors. Consequently, the centrosymmetric antiferromagnetic semiconductors -- in spite of their abundance in nature -- seem not promising for practical applications in spintronics~\cite{afmspin2,afmspin1,afmspin6,afmspin4,afmspin5,zunger2}.

Recently, efforts were made to explore the possible spin splittings hosted by \textit{all} types of magnetic space groups, involving non-magnetic, ferromagnetic, and antiferromagnetic materials (see, e.g., Refs.~\cite{zunger1,zunger2,zunger3,picozzibco,egorov2021,egorov2022,zeemanarxiv,zeemannpj}).  Several previously-overlooked spin-splitting patterns were discovered~\cite{zunger1,zunger2,zunger3,picozzibco,egorov2021,egorov2022,zeemanarxiv,zeemannpj}, but without demonstrating the possibility of creating and controlling sizable spin splittings in centrosymmetric antiferromagnets by magnetic or electric field. 
Interestingly, two works focusing on nonlinear photocurrent in MnBi$_2$Te$_4$~\cite{topologafm1} and magneto-optic Kerr effect in MnPSe$_3$~\cite{moke} (rather than spin splittings) hint to such a possibility. However, the general conditions and underlying mechanisms to the creation and control of spin splittings by magnetic or electric field in centrosymmetric antiferromagnets remain elusive.

In this Letter, we aim at exploring  spin splittings that would be controllable by electric field  and hosted by centrosymmetric antiferromagnets. Our basic idea is rooted in the magnetoelectric effect (see e.g., Ref.~\cite{mecoupling}). As a matter of fact, electric field not only creates polarization $P_\alpha$ but also generates magnetization $M_\beta \propto P_\alpha$ in magnetoelectric antiferromagnets ($\alpha,\beta=x,y,z$)~\cite{mecoupling}. The occurrence of $M_\beta$ implies an effective internal magnetic field $B^{\mathrm{eff}}_\beta \propto P_\alpha$ in materials, which couples with electronic spin $\sigma_\beta$ (i.e., Pauli matrix $\sigma_\beta$) and yields a Zeeman-like Hamiltonian $\lambda_{\alpha,\beta} P_\alpha \sigma_\beta$~\footnote{Note, however, that the coupling $\lambda_{\alpha,\beta}P_\alpha \sigma_\beta$ \textit{does not} exist in materials with time-reversal symmetry, as will be shown below.}.

Further, we check our idea by symmetry analysis and first-principles simulations. We identify twenty-one centrosymmetric antiferromagnetic point groups that accommodate electrically controllable Zeeman spin splittings. More promisingly, we find two centrosymmetric antiferromagnetic semiconductors, Fe$_2$TeO$_6$ and SrFe$_2$S$_2$O, in which large Zeeman spin splittings of $\sim$55 and $\sim$30 meV can be created by electric field of 6 MV/cm, respectively. The electronic spin magnetization associated to the splitting energy levels is confirmed to be switchable by reversing the electric field.\\

\begin{figure}[!h]
\includegraphics[width=0.95\linewidth]{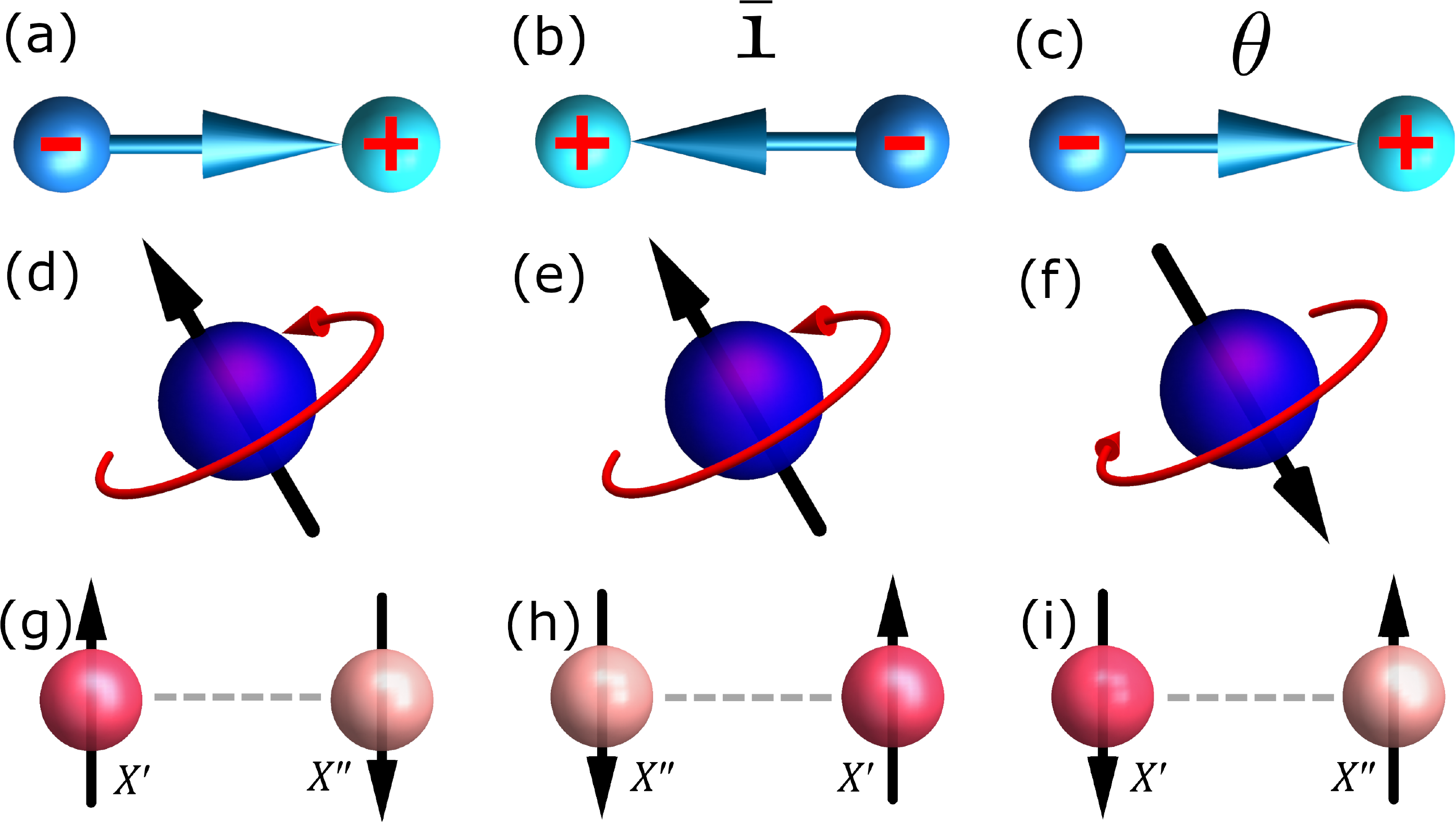}
\caption{\label{fig:spindispmag} The transformations of polarization $P_\alpha$ [panels (a)--(c)], spin angular momentum $S_\beta$ [panels (d)--(f)], and magnetic structure $X$ [panels (g)--(i)], under inversion $\bar{\mathfrak{1}}$ and time-reversal $\theta$. In (a)--(c), the ``$+$'' and ``$-$'' signs denote the positive and negative charges, respectively.  In (d)--(f), black arrow sketches the direction of $S_\beta$, while red curly arrow ``depicts'' spin. In (g)--(i), black arrow represents the magnetic moment carried by $X^\prime$ or $X^{\prime\prime}$ ions. The inversion center locates in the middle of the ``$+$'' and ``$-$'' spheres [panels (a)--(c)], at the center of the purple ball [panels (d)--(f)],  or at the midpoint of $X^\prime$ and $X^{\prime\prime}$ spheres [panels (g)--(i)], respectively.}
\end{figure}

\noindent
\textit{Couplings between polarization and spin. --} Among 122 magnetic point groups (MPGs), there are eleven groups, namely, $\bar{1}1^\prime$, $2/m1^\prime$, $mmm1^\prime$, $4/m1^\prime$, $4/mmm1^\prime$, $\bar{3}1^\prime$, $\bar{3}m1^\prime$, $6/m1^\prime$, $6/mmm1^\prime$, $m\bar{3}1^\prime$, and $m\bar{3}m1^\prime$, that contain both inversion $\bar{\mathfrak{1}}$ and time-reversal $\theta$ symmetries~\cite{mpoint,gtpack}. As will be shown below, these eleven groups -- belonging to type-II Shubnikov MPGs (denoted by $G^\prime$) -- host a sequence of subgroups (i.e., type-III Shubnikov MPGs) that allow the couplings between polarization and spin. Here, $G^\prime$ can be uniformly written as $G^\prime=G\cup\theta G$, where $G=G_0\cup \bar{\mathfrak{1}}G_0$ is the crystallographic point group and $G_0$ is the subgroup of $G$ containing \textit{only} proper rotations~\cite{gtpack,koster,character}.
We aim at finding the minimal couplings involving electric polarization and spin with respect to $G^\prime$ group.
To this end, we examine the transformation behaviors of electric polarization $P_\alpha$ and spin angular momentum operator $S_\beta$ under $\bar{\mathfrak{1}}$ and $\theta$, where $\alpha,\beta=x,y,z$ denote the Cartesian components (see Fig.~\ref{fig:spindispmag}).
Normally, the coupling $P_\alpha S_\beta$ (if existing) implies the electronic spin splittings induced by polarization $P_\alpha$, recalling that spin splitting is characterized by $\sigma_\beta$ where $S_\beta=\frac{\hbar}{2}\sigma_\beta$. 
Figures~\ref{fig:spindispmag}(a)--\ref{fig:spindispmag}(f) indicate the following transformation rules, namely, $\bar{\mathfrak{1}}: P_\alpha\rightarrow -P_\alpha, S_\beta\rightarrow S_\beta, \sigma_\beta \rightarrow \sigma_\beta$ and $\theta: P_\alpha\rightarrow P_\alpha, S_\beta\rightarrow -S_\beta, \sigma_\beta \rightarrow -\sigma_\beta$. Hence, the bilinear coupling between polarization and spin does not exist in the presence of either inversion or time-reversal symmetry, because $\bar{\mathfrak{1}}: P_\alpha \sigma_\beta \rightarrow -P_\alpha \sigma_\beta$; $\theta: P_\alpha \sigma_\beta \rightarrow -P_\alpha \sigma_\beta$.

We move on to explore whether the trilinear coupling $X P_\alpha \sigma_\beta$ does exist or not with respect to $G^\prime$. First, to fulfill the inversion and time-reversal symmetries, $X$ should be a quantity such that $\bar{\mathfrak{1}}: X\rightarrow -X$ and $\theta: X\rightarrow -X$.
Figures~\ref{fig:spindispmag}(g)--\ref{fig:spindispmag}(i) showcase such a possible $X$ extracted from an antiferromagnetic structure. 
For demonstrating purpose, we simply assume that the $X$ quantity, namely, magnetic order parameter, is formed by two atoms labelled by $X^\prime$ and $X^{\prime\prime}$, where $X^\prime$ and $X^{\prime\prime}$ are of the same atomic species, but carry magnetic moments along opposite directions [see Fig.~\ref{fig:spindispmag}(g)]. 
Under inversion $\bar{\mathfrak{1}}$, $X^\prime$ and $X^{\prime\prime}$ atoms swap their positions, while their carried magnetic moments remain unchanged [see Fig.~\ref{fig:spindispmag}(h)]; Under time-reversal $\theta$, $X^\prime$ and $X^{\prime\prime}$ atoms remain in place with the magnetic moments being flipped [see Fig.~\ref{fig:spindispmag}(i)]. This leads to $\bar{\mathfrak{1}}: X\rightarrow -X$ and $\theta: X\rightarrow -X$. Therefore, $X P_\alpha \sigma_\beta$ is compatible with inversion and time-reversal symmetries. 
Next, $X P_\alpha \sigma_\beta$ should be allowed by the proper rotation operations in $G_0$.
The $\bar{1}1^\prime$ is the simplest case to tackle, because its corresponding $G_0$ group only contains identity symmetry. Consequently, nine different couplings $X P_\alpha \sigma_\beta$ with $\alpha,\beta=x,y,z$ are permitted by symmetry operations of $\bar{1}1^\prime$ group. 
Unfortunately, the situation for the remaining ten type-II Shubnikov MPGs is quite complicated, since $G_0$ group contains more symmetry operations than identity, leading to additional symmetry constraint to $X P_\alpha \sigma_\beta$. 
For instance, $\mathfrak{2}_z$, rotation of $\pi$ along $z$ direction, transforms $P_x$, $P_z$, and $\sigma_z$ as $\mathfrak{2}_z: P_z\rightarrow P_z, P_x\rightarrow -P_x, \sigma_z\rightarrow \sigma_z$. 
As a result, $\mathfrak{2}_z$ transforms $X P_z \sigma_z$ and $X P_x \sigma_z$ via $\mathfrak{2}_z: X P_z \sigma_z \rightarrow X P_z \sigma_z, X P_x \sigma_z \rightarrow -X P_x \sigma_z$, assuming that $X$ is invariant under $\mathfrak{2}_z$. 
In this case, $X P_z \sigma_z$ is allowed by $\mathfrak{2}_z$ rotation, while $X P_x \sigma_z$ is not. Using this logic, we have conducted symmetry analysis regarding these ten type-II Shubnikov MPGs, shown in Section I of the Supplementary Material (SM)~\footnote{See Supplementary Material which includes symmetry analysis, methods, and some numerical results (e.g., band structures and orbital-projected spin magnetizations regarding SrFe$_2$S$_2$O and/or Fe$_2$TeO$_6$).} (containing Refs.~\cite{point,vasp1,vasp2,paw,ldaca,ldau,mma,materialsproject,materialsproject2,vesta,vaspkit,vaspkit2,matplotlib,seekpath,seekpath2,seekpath3,findsym1,findsym2,aroyo2006,aroyo20062,magndata,magndata2,magndata3,pyprocar,pyprocar2}). Taking $mmm1^\prime$ as an example [see Eq.~(S7) and Section I.3 of the SM], the symmetry-allowed trilinear couplings associated with $X \equiv M(A_{u})$ are given by $\lambda^\prime_{x,x} M(A_u) P_x \sigma_x + \lambda^\prime_{y,y} M(A_u) P_y \sigma_y +\lambda^\prime_{z,z} M(A_u) P_z \sigma_z$, where $\lambda^\prime_{x,x}$, $\lambda^\prime_{y,y}$, and $\lambda^\prime_{z,z}$ are coefficients characterizing the coupling strength. To understand the physical meaning of $X P_\alpha \sigma_\beta$ coupling, let us recall that $X P_\alpha \sigma_\beta$ is reminiscent of the conventional Zeeman term $B_\beta \sigma_\beta$~\cite{spintronic,rashbasoc}. This indicates that polarization $P_\alpha$ can generate in materials -- via a secondary effect -- an effective magnetic field $B^{\mathrm{eff}}_\beta \propto X P_\alpha$, whose microscopic origin may be roughly thought as follows: a polar distortion modifies electronic wave functions and ligand field in materials, yielding an internal effective magnetic field~\cite{mecoupling}. Such an effective field $B^{\mathrm{eff}}_\beta$ causes Zeeman spin splitting.

Now we demonstrate how to search for real materials  hosting $X P_\alpha \sigma_\beta$ coupling. First of all, note that the existence of order parameter $X$ breaks inversion $\bar{\mathfrak{1}}$, time-reversal $\theta$, and/or some other symmetry operations of $G^\prime$ group. 
Such symmetry breaking  lowers the symmetry of the system from $G^\prime$ group to its subgroup $g^\prime$ which contains the operations that are not broken by $X$. 
In such sense, $X$ is invariant under \textit{all} the symmetry operations of $g^\prime$. 
With respect to $g^\prime$ group, the effective Hamiltonian term $\lambda^\prime_{\alpha,\beta} X P_\alpha \sigma_\beta$ can be re-written as $\lambda_{\alpha,\beta} P_\alpha \sigma_\beta$, noting that the quantity $X$ is absorbed by the coefficient $\lambda^\prime_{\alpha,\beta}$. 
Therefore, to find a real material hosting $X$ order parameter and $\lambda_{\alpha,\beta} P_\alpha \sigma_\beta$ coupling, effort should be made to search for materials with magnetic point group $g^\prime$. Following this logic, we conduct symmetry analysis for the eleven aforementioned type-II Shubnikov MPGs, in order to extract the possible $g^\prime$ groups from $G^\prime$ (see Section I of the SM for the derivations). In particular, we find twenty-one type-III Shubnikov MPGs that accommodate the $\lambda_{\alpha,\beta} P_\alpha \sigma_\beta$ couplings, as summarized in Table~\ref{tab:pointmag} (see Section I.12 and Table S13 of the SM for more details). Interestingly, our derived Zeeman coupling coefficients (Table~\ref{tab:pointmag}) are similar to the tabulated magnetoelectric tensors~\cite{magnetoelectric}. In such sense, our proposed twenty-one MPGs also host the magnetoelectric effect, in agreement with our aforementioned analysis (see \textit{Introduction}). These MPGs do not have inversion $\bar{\mathfrak{1}}$ or time-reversal $\theta$, but rather exhibit parity-time symmetry  ($\bar{\mathfrak{1}}\theta$). In essence, these twenty-one type-III Shubnikov MPGs are centrosymmetric in the four-dimensional spacetime, since the $\bar{\mathfrak{1}}\theta$ symmetry operation transform the spatial-temporal coordinate $(x,y,z,t)$ to $(-x,-y,-z,-t)$. Hence, none of these twenty-one MPGs host spontaneous ferromagnetism or electric polarization. According to the $\lambda_{\alpha,\beta}P_\alpha \sigma_\beta$ coupling, the $\bar{\mathfrak{1}}\theta$ symmetry operation is broken in the presence of polarization, yielding Zeeman-type spin splittings. This coincides with the previous symmetry analysis which indicates that the breakdown of parity-time symmetry can generate spin splittings (see, e.g., Refs.~\cite{topologafm1,topologafm2,zunger1,zunger2,moke,kpmethod2009,grouptheorydresselhaus,zunger3}).

 Taking $m^\prime m^\prime m^\prime$ as an example, Table~\ref{tab:pointmag} indicates the $\lambda_{x,x}$, $\lambda_{y,y}$, and $\lambda_{z,z}$ couplings, yielding the effective Hamiltonian $H(m^\prime m^\prime m^\prime) = \lambda_{x,x} P_x \sigma_x + \lambda_{y,y} P_y \sigma_y + \lambda_{z,z} P_z \sigma_z =\kappa_{x,x} \mathcal{E}_x \sigma_x + \kappa_{y,y} \mathcal{E}_y \sigma_y + \kappa_{z,z} \mathcal{E}_z \sigma_z$, where $\mathcal{E}_\alpha$ is the electric field along the $\alpha$ direction~\footnote{Here, the second equality holds because that $\mathcal{E}_\alpha$ polarizes centrosymmetric materials by creating $P_\alpha$.}. Similarly, the effective Hamiltonians for $4/m^\prime m^\prime m^\prime$ and $\bar{3}^\prime m$ are given by $H(4/m^\prime m^\prime m^\prime)= \kappa_{x,x} (\mathcal{E}_x \sigma_x + \mathcal{E}_y \sigma_y)+\kappa_{z,z} \mathcal{E}_z \sigma_z$ and $H(\bar{3}^\prime m^\prime)=\kappa_{x,x}(\mathcal{E}_x\sigma_x+\mathcal{E}_y\sigma_y)+\kappa_{z,z}\mathcal{E}_z\sigma_z$, respectively. Note that the splittings predicted by $H(\bar{3}^\prime m^\prime)$ were claimed to be critical for the nonlinear photocurrent effect in topological material MnBi$_2$Te$_4$~\cite{topologafm1}.\\

\begin{table}[ht]
\caption{\label{tab:pointmag} The couplings that are hosted by twenty-one type-III Shubnikov MPGs. In each $(P_\alpha,\sigma_\beta)$ entry, $\lambda_{\alpha,\beta}$ indicates the coupling $\lambda_{\alpha,\beta}P_\alpha \sigma_\beta$; the ``$\ldots$'' implies that the coupling $P_\alpha \sigma_\beta$ is forbidden by symmetry. To better understand this table, we refer the readers to Section I.12 of the SM.} 
\begin{ruledtabular}
\begin{tabular}{cccccccccc}
  & \multicolumn{3}{c}{\cellcolor[HTML]{FE996B}$P_x$}                                     & \multicolumn{3}{c}{\cellcolor[HTML]{FFCC67}$P_y$}                                     & \multicolumn{3}{c}{\cellcolor[HTML]{38FFF8}$P_z$}                                     \\
 & \cellcolor[HTML]{FE996B}$\sigma_x$ & \cellcolor[HTML]{FE996B}$\sigma_y$ & \cellcolor[HTML]{FE996B}$\sigma_z$ & \cellcolor[HTML]{FFCC67}$\sigma_x$ & \cellcolor[HTML]{FFCC67}$\sigma_y$ & \cellcolor[HTML]{FFCC67}$\sigma_z$ & \cellcolor[HTML]{38FFF8}$\sigma_x$ & \cellcolor[HTML]{38FFF8}$\sigma_y$ & \cellcolor[HTML]{38FFF8}$\sigma_z$ \\
 \hline 
$\bar{1}^\prime$ & $\lambda_{x,x}$  & $\lambda_{x,y}$  & $\lambda_{x,z}$  & $\lambda_{y,x}$  & $\lambda_{y,y}$  & $\lambda_{y,z}$  & $\lambda_{z,x}$   & $\lambda_{z,y}$   & $\lambda_{z,z}$    \\  
$2/m^\prime$ & $\lambda_{x,x}$  & $\lambda_{x,y}$  & $\ldots$  & $\lambda_{y,x}$  & $\lambda_{y,y}$  & $\ldots$  & $\ldots$   & $\ldots$   & $\lambda_{z,z}$    \\  
$2^\prime/m$ & $\ldots$  & $\ldots$  & $\lambda_{x,z}$  & $\ldots$  & $\ldots$  & $\lambda_{y,z}$  & $\lambda_{z,x}$   & $\lambda_{z,y}$   & $\ldots$    \\ 
$m^\prime m^\prime m^\prime$ & $\lambda_{x,x}$  & $\ldots$  & $\ldots$  & $\ldots$  & $\lambda_{y,y}$  & $\ldots$  & $\ldots$   & $\ldots$   & $\lambda_{z,z}$    \\ 
$m m m^\prime$ & $\ldots$  & $\lambda_{x,y}$  & $\ldots$  & $\lambda_{y,x}$  & $\ldots$  & $\ldots$  & $\ldots$   & $\ldots$   & $\ldots$  \\ 
$m m^\prime m$ & $\ldots$  & $\ldots$  & $\lambda_{x,z}$  & $\ldots$  & $\ldots$  & $\ldots$  & $\lambda_{z,x}$   & $\ldots$   & $\ldots$  \\ 
$m^\prime m m$ & $\ldots$  & $\ldots$  & $\ldots$  & $\ldots$  & $\ldots$  & $\lambda_{y,z}$  & $\ldots$   & $\lambda_{z,y}$   & $\ldots$  \\ 
$4/m^\prime$ & $\lambda_{x,x}$  & $\lambda_{x,y}$  & $\ldots$  & $-\lambda_{x,y}$  & $\lambda_{x,x}$  & $\ldots$  & $\ldots$   & $\ldots$   & $\lambda_{z,z}$  \\
$4^\prime/m^\prime$ & $\lambda_{x,x}$  & $\lambda_{x,y}$  & $\ldots$  & $\lambda_{x,y}$  & $-\lambda_{x,x}$  & $\ldots$  & $\ldots$   & $\ldots$   & $\ldots$  \\
$4/m^\prime m^\prime m^\prime$ & $\lambda_{x,x}$  & $\ldots$  & $\ldots$  & $\ldots$  & $\lambda_{x,x}$  & $\ldots$  & $\ldots$   & $\ldots$   & $\lambda_{z,z}$  \\ 
$4/m^\prime m m$ & $\ldots$  &  $\lambda_{x,y}$  & $\ldots$  & $-\lambda_{x,y}$  & $\ldots$  & $\ldots$  & $\ldots$   & $\ldots$   & $\ldots$  \\ 
$4^\prime/m^\prime m^\prime m$ & $\lambda_{x,x}$  & $\ldots$  & $\ldots$  & $\ldots$  & $-\lambda_{x,x}$  & $\ldots$  & $\ldots$   & $\ldots$   & $\ldots$ \\
$4^\prime/m^\prime m m^\prime$ & $\ldots$  & $\lambda_{x,y}$  & $\ldots$  & $\lambda_{x,y}$  & $\ldots$  & $\ldots$  & $\ldots$   & $\ldots$   & $\ldots$ \\
$\bar{3}^\prime$ & $\lambda_{x,x}$  & $\lambda_{x,y}$  & $\ldots$  & $-\lambda_{x,y}$  & $\lambda_{x,x}$  & $\ldots$  & $\ldots$   & $\ldots$   & $\lambda_{z,z}$ \\
$\bar{3}^\prime m^\prime$ & $\lambda_{x,x}$  & $\ldots$  & $\ldots$  & $\ldots$  & $\lambda_{x,x}$  & $\ldots$  & $\ldots$   & $\ldots$   & $\lambda_{z,z}$ \\
$\bar{3}^\prime m$ & $\ldots$  & $\lambda_{x,y}$  & $\ldots$  & $-\lambda_{x,y}$  & $\ldots$  & $\ldots$  & $\ldots$   & $\ldots$   & $\ldots$ \\
$6/m^\prime$ & $\lambda_{x,x}$  & $\lambda_{x,y}$  & $\ldots$  & $-\lambda_{x,y}$  & $\lambda_{x,x}$  & $\ldots$  & $\ldots$   & $\ldots$   & $\lambda_{z,z}$  \\ 
$6/m^\prime m^\prime m^\prime$ & $\lambda_{x,x}$  & $\ldots$  & $\ldots$  & $\ldots$  & $\lambda_{x,x}$  & $\ldots$  & $\ldots$   & $\ldots$   & $\lambda_{z,z}$  \\ 
$6/m^\prime m m$ & $\ldots$  & $\lambda_{x,y}$  & $\ldots$  & $-\lambda_{x,y}$  & $\ldots$  & $\ldots$  & $\ldots$   & $\ldots$   & $\ldots$  \\
$m^\prime\bar{3}^\prime$ & $\lambda_{x,x}$  & $\ldots$  & $\ldots$  & $\ldots$  & $\lambda_{x,x}$  & $\ldots$  & $\ldots$   & $\ldots$   & $\lambda_{x,x}$  \\
$m^\prime\bar{3}^\prime m^\prime$ & $\lambda_{x,x}$  & $\ldots$  & $\ldots$  & $\ldots$  & $\lambda_{x,x}$  & $\ldots$  & $\ldots$   & $\ldots$   & $\lambda_{x,x}$  \\
\end{tabular}
\end{ruledtabular}
\end{table}

\begin{figure}[t]
\includegraphics[width=0.90\linewidth]{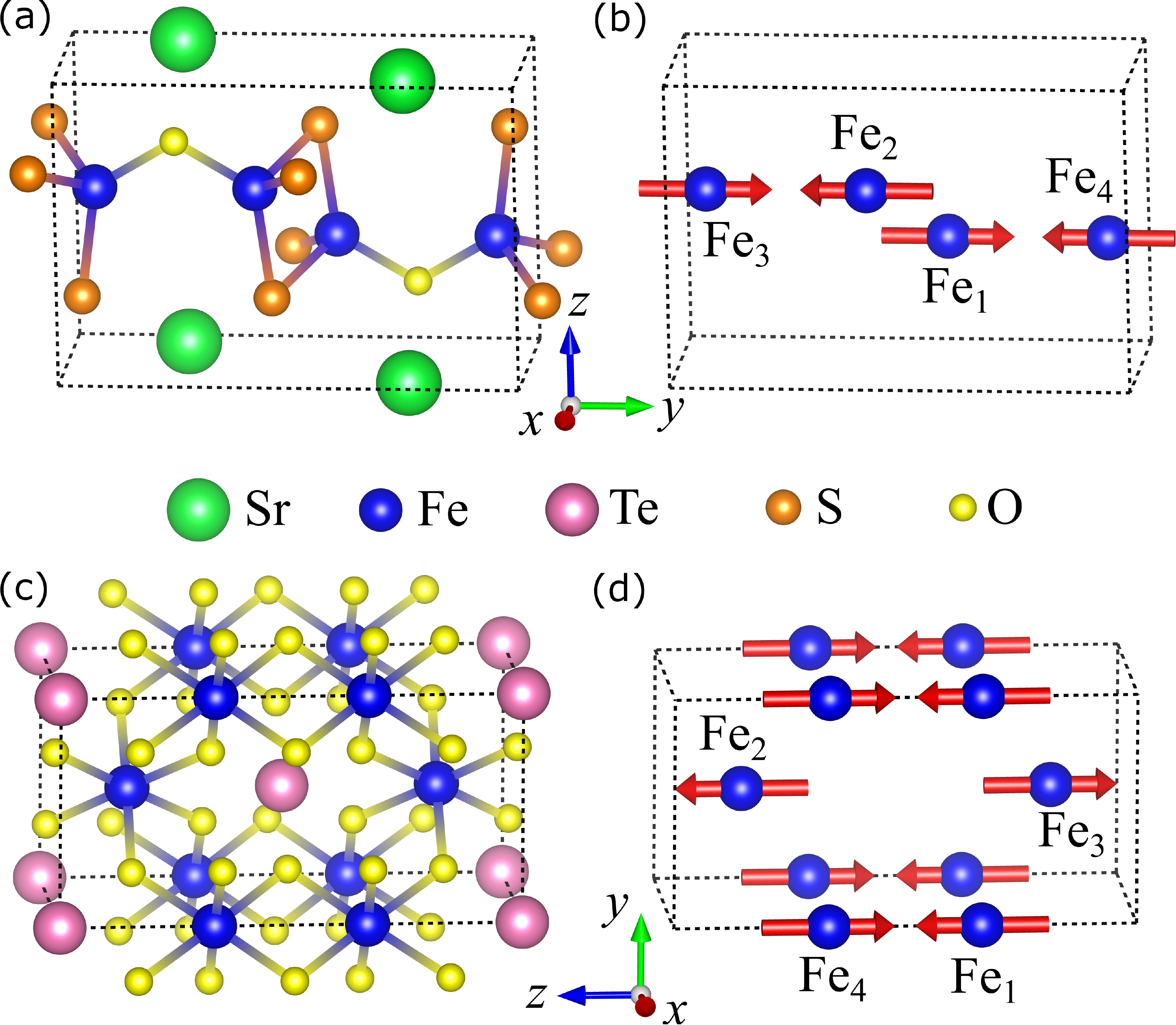}
\caption{\label{fig:crystalmag} Panels (a) and (c) show the crystal structures of SrFe$_2$S$_2$O and Fe$_2$TeO$_6$, respectively. Panels (b) and (d) sketch the ground state magnetic structures of SrFe$_2$S$_2$O (magnetic space group: $Pm^\prime m^\prime n^\prime$~\cite{magstr}) and Fe$_2$TeO$_6$ (magnetic space group: $P4_2/m^\prime n^\prime m^\prime$~\cite{fe2teo62}), respectively.}
\end{figure}

\begin{figure}[t]
\includegraphics[width=0.95\linewidth]{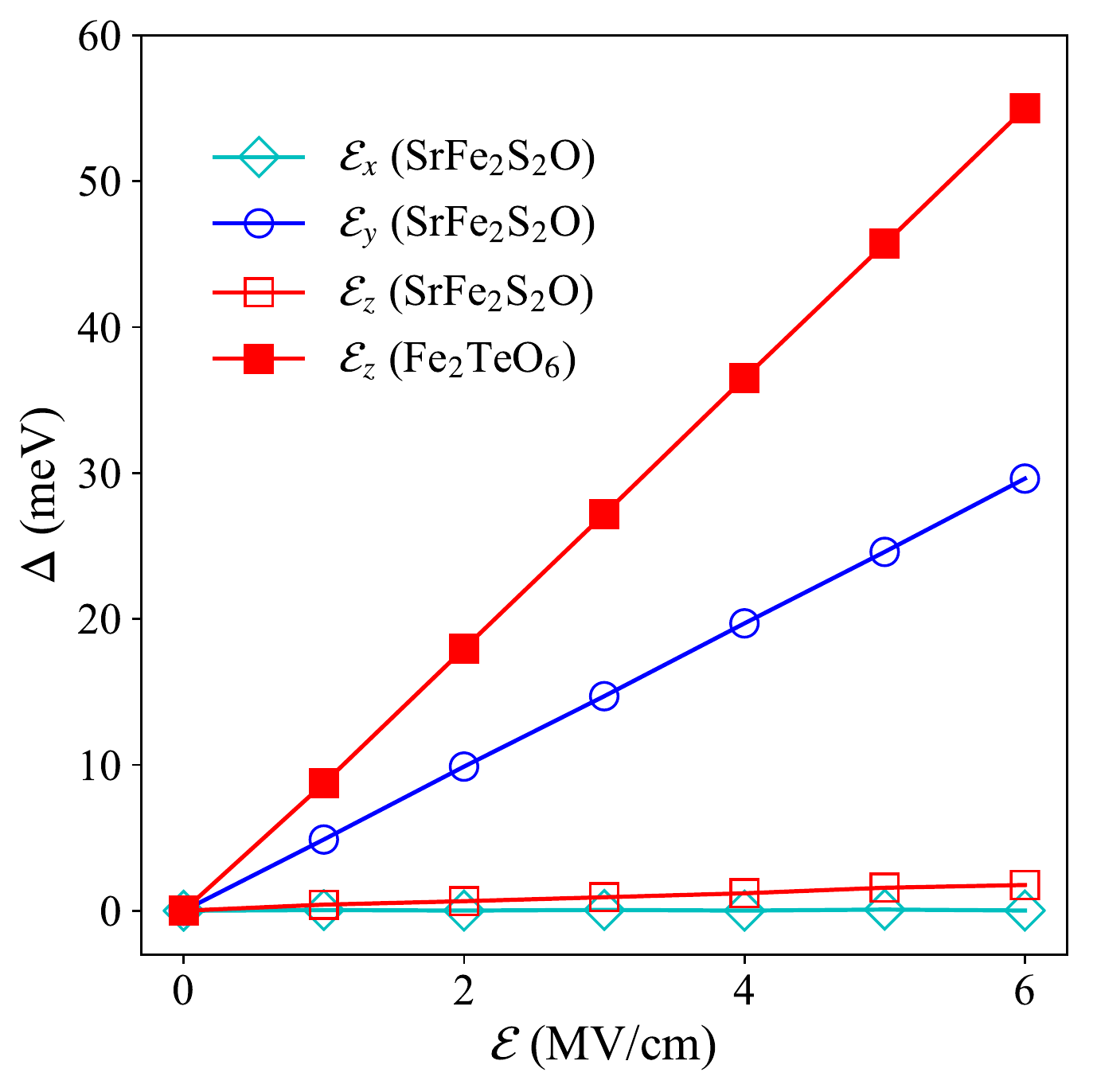}
\caption{\label{fig:Zeeman} The Zeeman spin splittings for the VBM of SrFe$_2$S$_2$O and the CBM of Fe$_2$TeO$_6$, as a function of electric field $\mathcal{E}$. We first determine the crystal structures (i.e., ionic degrees of freedom) of SrFe$_2$S$_2$O and Fe$_2$TeO$_6$ under each electric field, via a first-principles-based approach (see, e.g., Ref.~\cite{efield}). For each determined crystal structure, we then compute the energy levels and extract the Zeeman spin splitting at $\Gamma$ point. Note that during the calculation for energy levels, no elecric field is considered any more.}
\end{figure}

\noindent
\textit{Creating and controlling Zeeman splittings in SrFe$_2$S$_2$O and Fe$_2$TeO$_6$. --} Based on Table~\ref{tab:pointmag}, we search from the MAGNDATA database~\cite{magndata} for antiferromagnets with Zeeman splittings that can be created and controlled by electric field. Promisingly, we find two antiferromagnetic semiconductors, SrFe$_2$S$_2$O and Fe$_2$TeO$_6$ (see Fig.~\ref{fig:crystalmag}), whose N\'eel temperatures are both higher than 200 K~\cite{magstr,fe2teo6,fe2teo62}. The corresponding MPGs for SrFe$_2$S$_2$O and Fe$_2$TeO$_6$ are $m^\prime m^\prime m^\prime$~\cite{magstr} and $4/m^\prime m^\prime m^\prime$~\cite{fe2teo62}, respectively. Employing their ground state magnetic structures [sketched in Figs.~\ref{fig:crystalmag}(b) and~\ref{fig:crystalmag}(d)] and considering spin-orbit interaction, we use first-principles to compute the band structures of SrFe$_2$S$_2$O and Fe$_2$TeO$_6$ without polarization or with polarization created by electric field of 6 MV/cm (see Section III.1 of the SM). As shown in Fig.~S2 of the SM, the valence band maximum (VBM) of SrFe$_2$S$_2$O are located at the $\Gamma$ point, and the corresponding spin levels are (i) degenerate for non-polarized, (ii) nearly degenerate for $\mathcal{E}_x$-polarized, (iii) slightly split for $\mathcal{E}_z$-polarized, and (iv) obviously split for $\mathcal{E}_y$-polarized SrFe$_2$S$_2$O material. As for Fe$_2$TeO$_6$, the conduction band minimum (CBM) is at the $\Gamma$ point, and $\mathcal{E}_z$ apparently splits the spin levels at the CBM (see Fig.~S5 of the SM)~\footnote{We also numerically found that polarizing Fe$_2$TeO$_6$ by $\mathcal{E}_x=6$ MV/cm causes a tiny Zeeman spin splitting of $\sim$3 meV at the CBM.}. Our numerical simulations further indicate that the magnitudes of Zeeman spin splittings are in perfect linear relationship with $\mathcal{E}_\alpha$ (see Fig.~\ref{fig:Zeeman}). Strikingly, $\mathcal{E}_y=6$ MV/cm and $\mathcal{E}_z=6$ MV/cm generate Zeeman spin splittings of $\sim$30 meV and $\sim$55 meV~\footnote{The conventional Zeeman spin splitting created by magnetic field $B$ is given by $\lvert g^\prime \rvert \mu_B B$, where $g^\prime$ is the effective Land\'e $g$-factor~\cite{spintronic,spintronic2}. For free electrons or electrons in non-magnetic semiconductors with large enough band gap, the $g^\prime$ is nearly 2.0~\cite{spintronic,spintronic2}. The Zeeman spin splitting of $\sim$55 meV can be driven by magnetic field of $\sim$475 Tesla. In II-V wurtzite semiconductors ZnS and CdSe, the effective $\lvert g^\prime \rvert$ are 0.6 and 2.3, respectively~\cite{spintronic2}. Generating Zeeman splitting of $\sim$55 meV in ZnS and CdSe semiconductors thus requires a magnetic field of $\sim$413 and $\sim$1583 Tesla, respectively. In III-V semiconductor InSb, the effective $\lvert g^\prime \rvert$ can be as large as 51.3~\cite{spintronic2}. In such a case, magnetic field of $\sim$18 Tesla creates Zeeman spin splitting of $\sim$55 meV.}, respectively, for the VBM of SrFe$_2$S$_2$O and CBM of Fe$_2$TeO$_6$.

On the other hand, we notice that the spin splittings induced by $\mathcal{E}_x$, $\mathcal{E}_y$, and  $\mathcal{E}_z$ in SrFe$_2$S$_2$O exhibit highly distinct characteristics (see Fig.~\ref{fig:Zeeman}). For example, electric field $\mathcal{E}_x$ of 6 MV/cm~causes nearly null Zeeman spin splitting, implying the smallness of the coupling coefficient $\kappa_{x,x}$ in $H(m^\prime m^\prime m^\prime) =\kappa_{x,x} \mathcal{E}_x \sigma_x + \kappa_{y,y} \mathcal{E}_y \sigma_y + \kappa_{z,z} \mathcal{E}_z \sigma_z$. Meanwhile, the Zeeman spin splitting induced by $\mathcal{E}_y$ is far larger than that generated by $\mathcal{E}_z$ of the same magnitude as $\mathcal{E}_y$. 
For interpretation, we analyze the spin magnetization $(S_x,S_y,S_z)$ associated with the two top most energy sublevels at the $\Gamma$ point. 
When polarizing SrFe$_2$S$_2$O by $\mathcal{E}_\alpha$ ($\alpha=x,y,z$) electric field, an effective magnetic field $B^{\mathrm{eff}}_\alpha \propto \mathcal{E}_\alpha$ is created in the material. 
The $B^{\mathrm{eff}}_\alpha$ field couples with $S_\alpha$, causing a Zeeman energy proportional to $\pm B^{\mathrm{eff}}_\alpha S_\alpha$, where the $\pm$ sign characterizes the sublevels whose $\alpha$ spin magnetization component are positive or negative. 
Polarizing SrFe$_2$S$_2$O by $\mathcal{E}_y$=6 MV/cm~and $\mathcal{E}_z$=6 MV/cm~leads to the spin magnetization of $S_y\approx\pm0.74$ and $S_z\approx\pm0.08$, respectively. 
The predominant $S_y$ component implies that the Zeeman spin splitting created by $\mathcal{E}_y$ is the most prominent.
Our further analysis regarding the orbital-projected spin magnetization for SrFe$_2$S$_2$O can be found in Sections III.2 of the SM.

\begin{figure}[t]
\centering
\includegraphics[width=1.0\linewidth]{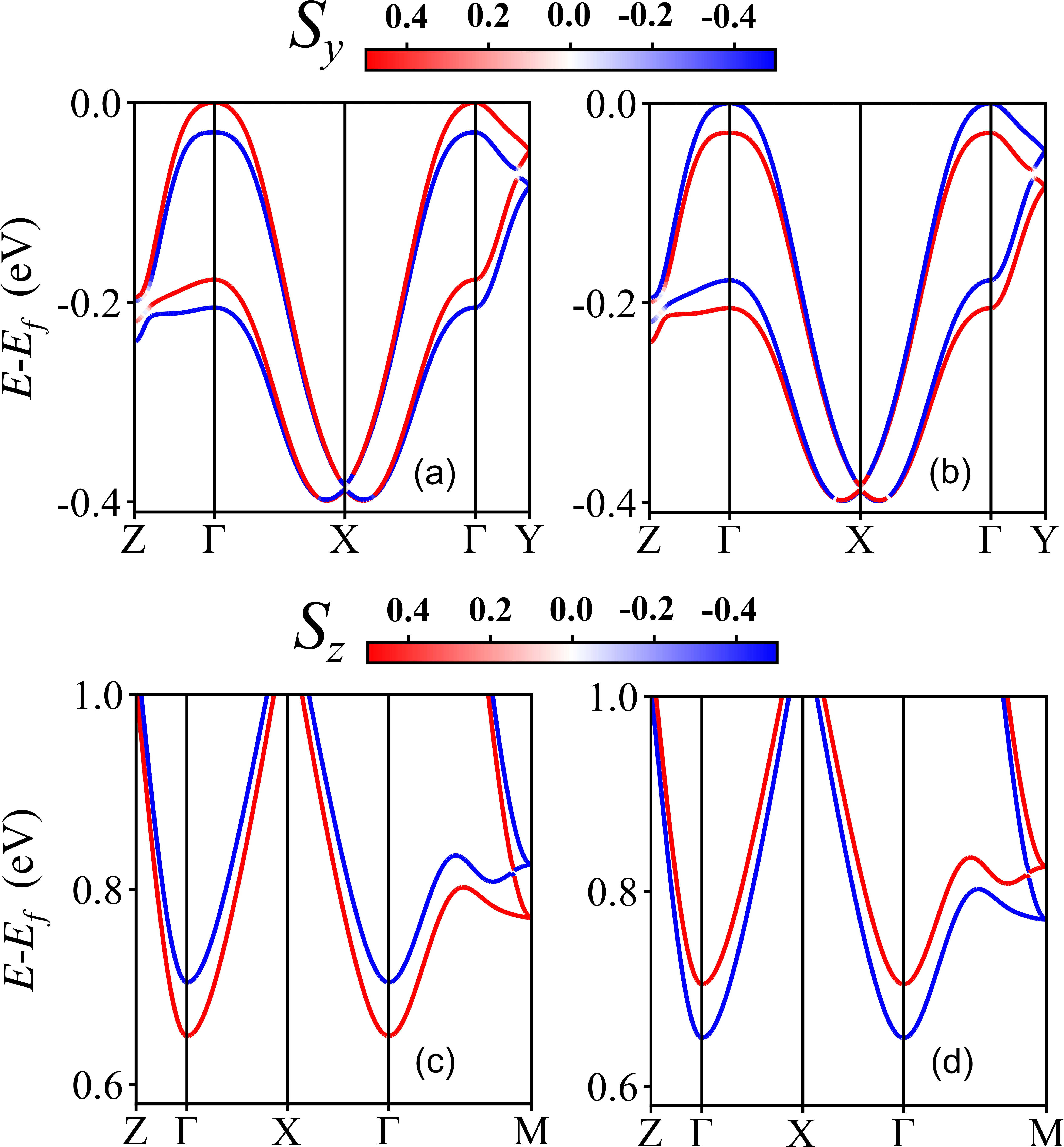}
\caption{\label{fig:ZeemanSFSOFTO} Panels (a) and (b) are local band structures of SrFe$_2$S$_2$O polarized by $\mathcal{E}_y=+6$ MV/cm~and $\mathcal{E}_y=-6$ MV/cm, respectively. Panels (c) and (d) show local band structures of Fe$_2$TeO$_6$ polarized by $\mathcal{E}_z=+6$ MV/cm~and $\mathcal{E}_z=-6$ MV/cm, respectively. The color bar corresponds to $S_y$ or $S_z$. The Fermi level $E_f$ is set as the VBM.}
\end{figure}

We now address whether the spin magnetization $S_\alpha$ for SrFe$_2$S$_2$O and Fe$_2$TeO$_6$ are switchable by electric field. To begin with, let us recall our model $H(m^\prime m^\prime m^\prime) = \kappa_{x,x} \mathcal{E}_x \sigma_x + \kappa_{y,y} \mathcal{E}_y \sigma_y + \kappa_{z,z} \mathcal{E}_z \sigma_z $ for SrFe$_2$S$_2$O ($m^\prime m^\prime m^\prime$ group). In the presence of $\mathcal{E}_y$, the spin levels will be split into two sublevels $E_{+}=\kappa_{y,y} \mathcal{E}_y$ (eigenstate being $\ket{+}$) and $E_{-}=-\kappa_{y,y} \mathcal{E}_y$ (eigenstate being $\ket{-}$), where $\sigma_y \ket{+}=\ket{+}$ and $\sigma_y \ket{-}=-\ket{-}$. The spin magnetization $S_y$ associated with $\kappa_{y,y} \mathcal{E}_y$ and $-\kappa_{y,y} \mathcal{E}_y$ are thus $\frac{1}{2}\bra{+}\sigma_y\ket{+}=\frac{1}{2}$ and $\frac{1}{2}\bra{-}\sigma_y\ket{-}=-\frac{1}{2}$~\footnote{Given the spinor quantum state $\ket{\psi}$, the expectation value of spin magnetization $S_\alpha$ is defined by $\frac{1}{2}\bra{\psi}\sigma_\alpha\ket{\psi}$ ($\alpha=x,y,z$). See, e.g., Ref.~\cite{rashbadresselhaus2}.}. When reversing electric field from $\mathcal{E}_y$ to $-\mathcal{E}_y$, the two split sublevels become $E_{-}=\kappa_{y,y} \mathcal{E}_y$ and $E_{+}=-\kappa_{y,y} \mathcal{E}_y$, with the corresponding eigenstates given by $\ket{-}$ and $\ket{+}$; Consequently, the $\kappa_{y,y} \mathcal{E}_y$ and $-\kappa_{y,y} \mathcal{E}_y$ sublevels are linked with the spin magnetization $S_y$ of $-\frac{1}{2}$ and $\frac{1}{2}$, respectively. Similarly, our models $H(m^\prime m^\prime m^\prime)$ and $H(4/m^\prime m^\prime m^\prime)= \kappa_{x,x} (\mathcal{E}_x \sigma_x + \mathcal{E}_y \sigma_y)+\kappa_{z,z} \mathcal{E}_z \sigma_z$ predicts that reversing electric field $\mathcal{E}_\alpha$ will switch the $S_\alpha$ spin magnetization between $\pm\frac{1}{2}$ and $\mp\frac{1}{2}$. To confirm our predictions, we compute the local band structures, along with spin magnetization $S_y$ or $S_z$, for SrFe$_2$S$_2$O and Fe$_2$TeO$_6$ (see Fig.~\ref{fig:ZeemanSFSOFTO}), including the spin-orbit interaction~\footnote{Neglecting the spin-orbit interaction does not qualitatively change our predictions for SrFe$_2$S$_2$O and Fe$_2$TeO$_6$ (see Fig.~S6 of the SM).}. We focus on the local bands around the VBM of SrFe$_2$S$_2$O and the CBM of Fe$_2$TeO$_6$. When reversing the electric field from $\mathcal{E}_y=+6$ MV/cm~to $\mathcal{E}_y=-6$ MV/cm (respectively, from $\mathcal{E}_z=+6$ MV/cm~to $\mathcal{E}_z=-6$ MV/cm), the $S_y$ for SrFe$_2$S$_2$O (respectively, $S_z$ for Fe$_2$TeO$_6$) is switchable. We further find that the $S_z$ component of SrFe$_2$S$_2$O can also be switched by $\mathcal{E}_z$, although the $S_z$ at the VBM is quite small (see Fig.~S4 of the SM)~\footnote{As for SrFe$_2$S$_2$O polarized by $\mathcal{E}_x=\pm6$  MV/cm, the induced spin splittings are too tiny to detect.}. Those results are in qualitative agreement with our model analysis.

To complete this section, let us comment on the limitation of our models. As mentioned above, our models predict $S_\alpha$ as $\pm\frac{1}{2}$ for SrFe$_2$S$_2$O, under electric field $\mathcal{E}_\alpha$. This is at odds with our first-principles simulations, which give, e.g., $S_y\approx\pm0.74$ and $S_z\approx\pm0.08$ for SrFe$_2$S$_2$O polarized by $\mathcal{E}_y$=6 MV/cm and $\mathcal{E}_z$=6 MV/cm, respectively. Note that the value $S_z=\pm\frac{1}{2}$ (predicted by our model) can be $\sim6$ times larger than the first-principles-predicted magnitudes (i.e., $S_z\approx\pm0.08$ of $\mathcal{E}_z$-polarized SrFe$_2$S$_2$O).
Such an inconsistency arises from the fact that that our models incorporate \textit{only} the minimal couplings involving electronic spin, electric field, and a mediated magnetic structure. Other degrees of freedom such as atomic orbitals and electronic wave vectors are neglected. More explicitly, our models consider merely two spin sublevels, while the first-principles calculations consider various degrees of freedom (e.g., the $3d$ orbitals of Fe ions), forming a multi-band case. As such, our models could only reveal the electric field induced Zeeman splittings qualitatively (i.e., not quantitatively).\\

\noindent
\textit{Summary and outlook. --} We have shown that electric field can create Zeeman spin splittings in centrosymmetric antiferromagnetic semiconductors belonging to one of the twenty-one MPGs (Table~\ref{tab:pointmag}). By first-principles simulations, we further identify two real materials, Fe$_2$TeO$_6$ and SrFe$_2$S$_2$O, that accommodate Zeeman spin splittings as large as $\sim$55 and $\sim$30 meV, respectively, in the presence of electric field of 6 MV/cm. The resulting Zeeman spin splittings are controllable by electric field, and can possibly be detected by some approaches (e.g., optical and transport measurements) that are well-established in spintronics~\cite{spintronic,spintronic2,spintronic3}. This will open a door towards the utilization of electronic spin in centrosymmetric antiferromagnetic semiconductors, emphasizing the importance of such materials for fabricating semiconductor spintronic devices. To finish, we hope that our discoveries will not only deepen the knowledge of magnetoelectric interactions, but also motivate a sequence of innovative studies in the emerging research directions of antiferromagnetic spintronics~\cite{afm,afmspin1,afmspin2,afmspin3,afmspin4,afmspin5,afmspin6} and semiconductor spintronics~\cite{rashbasoc,zeemanmagnet,spintronic,spintronic2,spintronic3}. \\

\noindent
\textit{Acknowledgements.--} This research was supported by the National Natural Science Foundation of China under Grants No. T2225013, No. 12274174, No. 12174142, No. 12034009, No. 11874207,  the Program for JLU Science and Technology Innovative Research Team, and the Science Challenge Project, No. TZ2016001. L. B. acknowledges the Vannevar Bush Faculty Fellowship (VBFF) Grant No. N00014-20-1-2834 from the Department of Defense
and the MonArk Quantum Foundry supported by the National Science Foundation Q-AMASE-i program under NSF Award No. DMR-1906383. We thank Prof. Y. Wei at Fudan University for the valuable discussion. The calculation was performed in the high-performance computing center of Jilin University.\\

\nolinenumbers


\begin{thebibliography}{73}%
\makeatletter
\providecommand \@ifxundefined [1]{%
 \@ifx{#1\undefined}
}%
\providecommand \@ifnum [1]{%
 \ifnum #1\expandafter \@firstoftwo
 \else \expandafter \@secondoftwo
 \fi
}%
\providecommand \@ifx [1]{%
 \ifx #1\expandafter \@firstoftwo
 \else \expandafter \@secondoftwo
 \fi
}%
\providecommand \natexlab [1]{#1}%
\providecommand \enquote  [1]{``#1''}%
\providecommand \bibnamefont  [1]{#1}%
\providecommand \bibfnamefont [1]{#1}%
\providecommand \citenamefont [1]{#1}%
\providecommand \href@noop [0]{\@secondoftwo}%
\providecommand \href [0]{\begingroup \@sanitize@url \@href}%
\providecommand \@href[1]{\@@startlink{#1}\@@href}%
\providecommand \@@href[1]{\endgroup#1\@@endlink}%
\providecommand \@sanitize@url [0]{\catcode `\\12\catcode `\$12\catcode
  `\&12\catcode `\#12\catcode `\^12\catcode `\_12\catcode `\%12\relax}%
\providecommand \@@startlink[1]{}%
\providecommand \@@endlink[0]{}%
\providecommand \url  [0]{\begingroup\@sanitize@url \@url }%
\providecommand \@url [1]{\endgroup\@href {#1}{\urlprefix }}%
\providecommand \urlprefix  [0]{URL }%
\providecommand \Eprint [0]{\href }%
\providecommand \doibase [0]{http://dx.doi.org/}%
\providecommand \selectlanguage [0]{\@gobble}%
\providecommand \bibinfo  [0]{\@secondoftwo}%
\providecommand \bibfield  [0]{\@secondoftwo}%
\providecommand \translation [1]{[#1]}%
\providecommand \BibitemOpen [0]{}%
\providecommand \bibitemStop [0]{}%
\providecommand \bibitemNoStop [0]{.\EOS\space}%
\providecommand \EOS [0]{\spacefactor3000\relax}%
\providecommand \BibitemShut  [1]{\csname bibitem#1\endcsname}%
\let\auto@bib@innerbib\@empty
\bibitem [{\citenamefont {Sch\"{a}pers}(2016)}]{spintronic}%
  \BibitemOpen
  \bibfield  {author} {\bibinfo {author} {\bibfnamefont {T.}~\bibnamefont
  {Sch\"{a}pers}},\ }\href@noop {} {\emph {\bibinfo {title} {Semiconductor
  Spintronics}}}\ (\bibinfo  {publisher} {De Gruyter},\ \bibinfo {year}
  {2016})\BibitemShut {NoStop}%
\bibitem [{\citenamefont {Xia}\ \emph {et~al.}(2011)\citenamefont {Xia},
  \citenamefont {Ge},\ and\ \citenamefont {Chang}}]{spintronic2}%
  \BibitemOpen
  \bibfield  {author} {\bibinfo {author} {\bibfnamefont {J.}~\bibnamefont
  {Xia}}, \bibinfo {author} {\bibfnamefont {W.}~\bibnamefont {Ge}}, \ and\
  \bibinfo {author} {\bibfnamefont {K.}~\bibnamefont {Chang}},\ }\href@noop {}
  {\emph {\bibinfo {title} {Semiconductor Spintronics}}}\ (\bibinfo
  {publisher} {{WORLD} {SCIENTIFIC}},\ \bibinfo {year} {2011})\BibitemShut
  {NoStop}%
\bibitem [{\citenamefont {Wang}\ \emph {et~al.}(2020)\citenamefont {Wang},
  \citenamefont {Gopal}, \citenamefont {Picozzi}, \citenamefont {Curtarolo},
  \citenamefont {Nardelli},\ and\ \citenamefont
  {S{\l}awi{\'{n}}ska}}]{spinhall}%
  \BibitemOpen
  \bibfield  {author} {\bibinfo {author} {\bibfnamefont {H.}~\bibnamefont
  {Wang}}, \bibinfo {author} {\bibfnamefont {P.}~\bibnamefont {Gopal}},
  \bibinfo {author} {\bibfnamefont {S.}~\bibnamefont {Picozzi}}, \bibinfo
  {author} {\bibfnamefont {S.}~\bibnamefont {Curtarolo}}, \bibinfo {author}
  {\bibfnamefont {M.~B.}\ \bibnamefont {Nardelli}}, \ and\ \bibinfo {author}
  {\bibfnamefont {J.}~\bibnamefont {S{\l}awi{\'{n}}ska}},\ }\href@noop {}
  {\bibfield  {journal} {\bibinfo  {journal} {npj Comput. Mater.}\ }\textbf
  {\bibinfo {volume} {6}},\ \bibinfo {pages} {7} (\bibinfo {year}
  {2020})}\BibitemShut {NoStop}%
\bibitem [{\citenamefont {Picozzi}(2014)}]{rashbadresselhaus1}%
  \BibitemOpen
  \bibfield  {author} {\bibinfo {author} {\bibfnamefont {S.}~\bibnamefont
  {Picozzi}},\ }\href {\doibase 10.3389/fphy.2014.00010} {\bibfield  {journal}
  {\bibinfo  {journal} {Front. Phys.}\ }\textbf {\bibinfo {volume} {2}},\
  \bibinfo {pages} {10} (\bibinfo {year} {2014})}\BibitemShut {NoStop}%
\bibitem [{\citenamefont {Tao}\ and\ \citenamefont
  {Tsymbal}(2021)}]{rashbadresselhaus2}%
  \BibitemOpen
  \bibfield  {author} {\bibinfo {author} {\bibfnamefont {L.~L.}\ \bibnamefont
  {Tao}}\ and\ \bibinfo {author} {\bibfnamefont {E.~Y.}\ \bibnamefont
  {Tsymbal}},\ }\href {\doibase 10.1088/1361-6463/abcc25} {\bibfield  {journal}
  {\bibinfo  {journal} {J. Phys. D: Appl. Phys.}\ }\textbf {\bibinfo {volume}
  {54}},\ \bibinfo {pages} {113001} (\bibinfo {year} {2021})}\BibitemShut
  {NoStop}%
\bibitem [{\citenamefont {Bychkov}\ and\ \citenamefont
  {Rashba}(1984)}]{rashbajetp}%
  \BibitemOpen
  \bibfield  {author} {\bibinfo {author} {\bibfnamefont {Y.~A.}\ \bibnamefont
  {Bychkov}}\ and\ \bibinfo {author} {\bibfnamefont {E.~I.}\ \bibnamefont
  {Rashba}},\ }\href@noop {} {\bibfield  {journal} {\bibinfo  {journal} {JETP
  Lett.}\ }\textbf {\bibinfo {volume} {39}},\ \bibinfo {pages} {78} (\bibinfo
  {year} {1984})}\BibitemShut {NoStop}%
\bibitem [{\citenamefont {Dresselhaus}(1955)}]{dresselhausoriginal}%
  \BibitemOpen
  \bibfield  {author} {\bibinfo {author} {\bibfnamefont {G.}~\bibnamefont
  {Dresselhaus}},\ }\href@noop {} {\bibfield  {journal} {\bibinfo  {journal}
  {Phys. Rev.}\ }\textbf {\bibinfo {volume} {100}},\ \bibinfo {pages} {580}
  (\bibinfo {year} {1955})}\BibitemShut {NoStop}%
\bibitem [{\citenamefont {Wolf}\ \emph {et~al.}(2001)\citenamefont {Wolf},
  \citenamefont {Awschalom}, \citenamefont {Buhrman}, \citenamefont {Daughton},
  \citenamefont {von Moln\'ar}, \citenamefont {Roukes}, \citenamefont
  {Chtchelkanova},\ and\ \citenamefont {Treger}}]{zeemanmagnet}%
  \BibitemOpen
  \bibfield  {author} {\bibinfo {author} {\bibfnamefont {S.~A.}\ \bibnamefont
  {Wolf}}, \bibinfo {author} {\bibfnamefont {D.~D.}\ \bibnamefont {Awschalom}},
  \bibinfo {author} {\bibfnamefont {R.~A.}\ \bibnamefont {Buhrman}}, \bibinfo
  {author} {\bibfnamefont {J.~M.}\ \bibnamefont {Daughton}}, \bibinfo {author}
  {\bibfnamefont {S.}~\bibnamefont {von Moln\'ar}}, \bibinfo {author}
  {\bibfnamefont {M.~L.}\ \bibnamefont {Roukes}}, \bibinfo {author}
  {\bibfnamefont {A.~Y.}\ \bibnamefont {Chtchelkanova}}, \ and\ \bibinfo
  {author} {\bibfnamefont {D.~M.}\ \bibnamefont {Treger}},\ }\href@noop {}
  {\bibfield  {journal} {\bibinfo  {journal} {Science}\ }\textbf {\bibinfo
  {volume} {294}},\ \bibinfo {pages} {1488} (\bibinfo {year}
  {2001})}\BibitemShut {NoStop}%
\bibitem [{\citenamefont {N{\v{e}}mec}\ \emph {et~al.}(2018)\citenamefont
  {N{\v{e}}mec}, \citenamefont {Fiebig}, \citenamefont {Kampfrath},\ and\
  \citenamefont {Kimel}}]{afmspin2}%
  \BibitemOpen
  \bibfield  {author} {\bibinfo {author} {\bibfnamefont {P.}~\bibnamefont
  {N{\v{e}}mec}}, \bibinfo {author} {\bibfnamefont {M.}~\bibnamefont {Fiebig}},
  \bibinfo {author} {\bibfnamefont {T.}~\bibnamefont {Kampfrath}}, \ and\
  \bibinfo {author} {\bibfnamefont {A.~V.}\ \bibnamefont {Kimel}},\ }\href
  {\doibase 10.1038/s41567-018-0051-x} {\bibfield  {journal} {\bibinfo
  {journal} {Nat. Phys.}\ }\textbf {\bibinfo {volume} {14}},\ \bibinfo {pages}
  {229} (\bibinfo {year} {2018})}\BibitemShut {NoStop}%
\bibitem [{\citenamefont {{\v{Z}}elezn{\'{y}}}\ \emph
  {et~al.}(2018)\citenamefont {{\v{Z}}elezn{\'{y}}}, \citenamefont {Wadley},
  \citenamefont {Olejn{\'{\i}}k}, \citenamefont {Hoffmann},\ and\ \citenamefont
  {Ohno}}]{afmspin1}%
  \BibitemOpen
  \bibfield  {author} {\bibinfo {author} {\bibfnamefont {J.}~\bibnamefont
  {{\v{Z}}elezn{\'{y}}}}, \bibinfo {author} {\bibfnamefont {P.}~\bibnamefont
  {Wadley}}, \bibinfo {author} {\bibfnamefont {K.}~\bibnamefont
  {Olejn{\'{\i}}k}}, \bibinfo {author} {\bibfnamefont {A.}~\bibnamefont
  {Hoffmann}}, \ and\ \bibinfo {author} {\bibfnamefont {H.}~\bibnamefont
  {Ohno}},\ }\href {\doibase 10.1038/s41567-018-0062-7} {\bibfield  {journal}
  {\bibinfo  {journal} {Nat. Phys.}\ }\textbf {\bibinfo {volume} {14}},\
  \bibinfo {pages} {220} (\bibinfo {year} {2018})}\BibitemShut {NoStop}%
\bibitem [{\citenamefont {Fukami}\ \emph {et~al.}(2020)\citenamefont {Fukami},
  \citenamefont {Lorenz},\ and\ \citenamefont {Gomonay}}]{afmspin6}%
  \BibitemOpen
  \bibfield  {author} {\bibinfo {author} {\bibfnamefont {S.}~\bibnamefont
  {Fukami}}, \bibinfo {author} {\bibfnamefont {V.~O.}\ \bibnamefont {Lorenz}},
  \ and\ \bibinfo {author} {\bibfnamefont {O.}~\bibnamefont {Gomonay}},\ }\href
  {\doibase 10.1063/5.0023614} {\bibfield  {journal} {\bibinfo  {journal} {J.
  Appl. Phys.}\ }\textbf {\bibinfo {volume} {128}},\ \bibinfo {pages} {070401}
  (\bibinfo {year} {2020})}\BibitemShut {NoStop}%
\bibitem [{\citenamefont {Jungwirth}\ \emph {et~al.}(2018)\citenamefont
  {Jungwirth}, \citenamefont {Sinova}, \citenamefont {Manchon}, \citenamefont
  {Marti}, \citenamefont {Wunderlich},\ and\ \citenamefont
  {Felser}}]{afmspin4}%
  \BibitemOpen
  \bibfield  {author} {\bibinfo {author} {\bibfnamefont {T.}~\bibnamefont
  {Jungwirth}}, \bibinfo {author} {\bibfnamefont {J.}~\bibnamefont {Sinova}},
  \bibinfo {author} {\bibfnamefont {A.}~\bibnamefont {Manchon}}, \bibinfo
  {author} {\bibfnamefont {X.}~\bibnamefont {Marti}}, \bibinfo {author}
  {\bibfnamefont {J.}~\bibnamefont {Wunderlich}}, \ and\ \bibinfo {author}
  {\bibfnamefont {C.}~\bibnamefont {Felser}},\ }\href {\doibase
  10.1038/s41567-018-0063-6} {\bibfield  {journal} {\bibinfo  {journal} {Nat.
  Phys.}\ }\textbf {\bibinfo {volume} {14}},\ \bibinfo {pages} {200} (\bibinfo
  {year} {2018})}\BibitemShut {NoStop}%
\bibitem [{\citenamefont {Jungwirth}\ \emph {et~al.}(2016)\citenamefont
  {Jungwirth}, \citenamefont {Marti}, \citenamefont {Wadley},\ and\
  \citenamefont {Wunderlich}}]{afmspin5}%
  \BibitemOpen
  \bibfield  {author} {\bibinfo {author} {\bibfnamefont {T.}~\bibnamefont
  {Jungwirth}}, \bibinfo {author} {\bibfnamefont {X.}~\bibnamefont {Marti}},
  \bibinfo {author} {\bibfnamefont {P.}~\bibnamefont {Wadley}}, \ and\ \bibinfo
  {author} {\bibfnamefont {J.}~\bibnamefont {Wunderlich}},\ }\href {\doibase
  10.1038/nnano.2016.18} {\bibfield  {journal} {\bibinfo  {journal} {Nat.
  Nanotechnol.}\ }\textbf {\bibinfo {volume} {11}},\ \bibinfo {pages} {231}
  (\bibinfo {year} {2016})}\BibitemShut {NoStop}%
\bibitem [{\citenamefont {Yuan}\ \emph
  {et~al.}(2021{\natexlab{a}})\citenamefont {Yuan}, \citenamefont {Wang},
  \citenamefont {Luo},\ and\ \citenamefont {Zunger}}]{zunger2}%
  \BibitemOpen
  \bibfield  {author} {\bibinfo {author} {\bibfnamefont {L.-D.}\ \bibnamefont
  {Yuan}}, \bibinfo {author} {\bibfnamefont {Z.}~\bibnamefont {Wang}}, \bibinfo
  {author} {\bibfnamefont {J.-W.}\ \bibnamefont {Luo}}, \ and\ \bibinfo
  {author} {\bibfnamefont {A.}~\bibnamefont {Zunger}},\ }\href {\doibase
  10.1103/PhysRevMaterials.5.014409} {\bibfield  {journal} {\bibinfo  {journal}
  {Phys. Rev. Materials}\ }\textbf {\bibinfo {volume} {5}},\ \bibinfo {pages}
  {014409} (\bibinfo {year} {2021}{\natexlab{a}})}\BibitemShut {NoStop}%
\bibitem [{\citenamefont {Yuan}\ \emph
  {et~al.}(2021{\natexlab{b}})\citenamefont {Yuan}, \citenamefont {Wang},
  \citenamefont {Luo},\ and\ \citenamefont {Zunger}}]{zunger1}%
  \BibitemOpen
  \bibfield  {author} {\bibinfo {author} {\bibfnamefont {L.-D.}\ \bibnamefont
  {Yuan}}, \bibinfo {author} {\bibfnamefont {Z.}~\bibnamefont {Wang}}, \bibinfo
  {author} {\bibfnamefont {J.-W.}\ \bibnamefont {Luo}}, \ and\ \bibinfo
  {author} {\bibfnamefont {A.}~\bibnamefont {Zunger}},\ }\href {\doibase
  10.1103/PhysRevB.103.224410} {\bibfield  {journal} {\bibinfo  {journal}
  {Phys. Rev. B}\ }\textbf {\bibinfo {volume} {103}},\ \bibinfo {pages}
  {224410} (\bibinfo {year} {2021}{\natexlab{b}})}\BibitemShut {NoStop}%
\bibitem [{\citenamefont {Yuan}\ \emph {et~al.}(2020)\citenamefont {Yuan},
  \citenamefont {Wang}, \citenamefont {Luo}, \citenamefont {Rashba},\ and\
  \citenamefont {Zunger}}]{zunger3}%
  \BibitemOpen
  \bibfield  {author} {\bibinfo {author} {\bibfnamefont {L.-D.}\ \bibnamefont
  {Yuan}}, \bibinfo {author} {\bibfnamefont {Z.}~\bibnamefont {Wang}}, \bibinfo
  {author} {\bibfnamefont {J.-W.}\ \bibnamefont {Luo}}, \bibinfo {author}
  {\bibfnamefont {E.~I.}\ \bibnamefont {Rashba}}, \ and\ \bibinfo {author}
  {\bibfnamefont {A.}~\bibnamefont {Zunger}},\ }\href {\doibase
  10.1103/PhysRevB.102.014422} {\bibfield  {journal} {\bibinfo  {journal}
  {Phys. Rev. B}\ }\textbf {\bibinfo {volume} {102}},\ \bibinfo {pages}
  {014422} (\bibinfo {year} {2020})}\BibitemShut {NoStop}%
\bibitem [{\citenamefont {Yamauchi}\ \emph {et~al.}(2019)\citenamefont
  {Yamauchi}, \citenamefont {Barone},\ and\ \citenamefont
  {Picozzi}}]{picozzibco}%
  \BibitemOpen
  \bibfield  {author} {\bibinfo {author} {\bibfnamefont {K.}~\bibnamefont
  {Yamauchi}}, \bibinfo {author} {\bibfnamefont {P.}~\bibnamefont {Barone}}, \
  and\ \bibinfo {author} {\bibfnamefont {S.}~\bibnamefont {Picozzi}},\
  }\href@noop {} {\bibfield  {journal} {\bibinfo  {journal} {Phys. Rev. B}\
  }\textbf {\bibinfo {volume} {100}},\ \bibinfo {pages} {245115} (\bibinfo
  {year} {2019})}\BibitemShut {NoStop}%
\bibitem [{\citenamefont {Egorov}\ \emph {et~al.}(2021)\citenamefont {Egorov},
  \citenamefont {Litvin},\ and\ \citenamefont {Evarestov}}]{egorov2021}%
  \BibitemOpen
  \bibfield  {author} {\bibinfo {author} {\bibfnamefont {S.~A.}\ \bibnamefont
  {Egorov}}, \bibinfo {author} {\bibfnamefont {D.~B.}\ \bibnamefont {Litvin}},
  \ and\ \bibinfo {author} {\bibfnamefont {R.~A.}\ \bibnamefont {Evarestov}},\
  }\href@noop {} {\bibfield  {journal} {\bibinfo  {journal} {J. Phys. Chem. C}\
  }\textbf {\bibinfo {volume} {125}},\ \bibinfo {pages} {16147} (\bibinfo
  {year} {2021})}\BibitemShut {NoStop}%
\bibitem [{\citenamefont {Egorov}\ and\ \citenamefont
  {Evarestov}(2022)}]{egorov2022}%
  \BibitemOpen
  \bibfield  {author} {\bibinfo {author} {\bibfnamefont {S.~A.}\ \bibnamefont
  {Egorov}}\ and\ \bibinfo {author} {\bibfnamefont {R.~A.}\ \bibnamefont
  {Evarestov}},\ }\href {\doibase 10.1016/j.physe.2021.115118} {\bibfield
  {journal} {\bibinfo  {journal} {Physica E: Low-dimens. Syst. Nanostruct.}\
  }\textbf {\bibinfo {volume} {139}},\ \bibinfo {pages} {115118} (\bibinfo
  {year} {2022})}\BibitemShut {NoStop}%
\bibitem [{\citenamefont {Reichlov\'{a}}\ \emph {et~al.}(2021)\citenamefont
  {Reichlov\'{a}}, \citenamefont {Seeger}, \citenamefont
  {Gonz\'{a}lez-Hern\'{a}ndez}, \citenamefont {Kounta}, \citenamefont
  {Schlitz}, \citenamefont {Kriegner}, \citenamefont {Ritzinger}, \citenamefont
  {Lammel}, \citenamefont {Leivisk\"{a}}, \citenamefont {Pet\v{r}\'{i}\v{c}ek},
  \citenamefont {Dole\v{z}al}, \citenamefont {Schmoranzerov\'{a}},
  \citenamefont {Bad'ura}, \citenamefont {Thomas}, \citenamefont {Baltz},
  \citenamefont {Michez}, \citenamefont {Sinova}, \citenamefont {Goennenwein},
  \citenamefont {Jungwirth},\ and\ \citenamefont {\v{S}mejkal}}]{zeemanarxiv}%
  \BibitemOpen
  \bibfield  {author} {\bibinfo {author} {\bibfnamefont {H.}~\bibnamefont
  {Reichlov\'{a}}}, \bibinfo {author} {\bibfnamefont {R.~L.}\ \bibnamefont
  {Seeger}}, \bibinfo {author} {\bibfnamefont {R.}~\bibnamefont
  {Gonz\'{a}lez-Hern\'{a}ndez}}, \bibinfo {author} {\bibfnamefont
  {I.}~\bibnamefont {Kounta}}, \bibinfo {author} {\bibfnamefont
  {R.}~\bibnamefont {Schlitz}}, \bibinfo {author} {\bibfnamefont
  {D.}~\bibnamefont {Kriegner}}, \bibinfo {author} {\bibfnamefont
  {P.}~\bibnamefont {Ritzinger}}, \bibinfo {author} {\bibfnamefont
  {M.}~\bibnamefont {Lammel}}, \bibinfo {author} {\bibfnamefont
  {M.}~\bibnamefont {Leivisk\"{a}}}, \bibinfo {author} {\bibfnamefont
  {V.}~\bibnamefont {Pet\v{r}\'{i}\v{c}ek}}, \bibinfo {author} {\bibfnamefont
  {P.}~\bibnamefont {Dole\v{z}al}}, \bibinfo {author} {\bibfnamefont
  {E.}~\bibnamefont {Schmoranzerov\'{a}}}, \bibinfo {author} {\bibfnamefont
  {A.}~\bibnamefont {Bad'ura}}, \bibinfo {author} {\bibfnamefont
  {A.}~\bibnamefont {Thomas}}, \bibinfo {author} {\bibfnamefont
  {V.}~\bibnamefont {Baltz}}, \bibinfo {author} {\bibfnamefont
  {L.}~\bibnamefont {Michez}}, \bibinfo {author} {\bibfnamefont
  {J.}~\bibnamefont {Sinova}}, \bibinfo {author} {\bibfnamefont {S.~T.~B.}\
  \bibnamefont {Goennenwein}}, \bibinfo {author} {\bibfnamefont
  {T.}~\bibnamefont {Jungwirth}}, \ and\ \bibinfo {author} {\bibfnamefont
  {L.}~\bibnamefont {\v{S}mejkal}},\ }\href@noop {} {\enquote {\bibinfo {title}
  {Macroscopic time reversal symmetry breaking by staggered spin-momentum
  interaction},}\ } (\bibinfo {year} {2021}),\ \Eprint
  {http://arxiv.org/abs/arXiv:2012.15651} {arXiv:2012.15651} \BibitemShut
  {NoStop}%
\bibitem [{\citenamefont {Ramazashvili}\ \emph {et~al.}(2021)\citenamefont
  {Ramazashvili}, \citenamefont {Grigoriev}, \citenamefont {Helm},
  \citenamefont {Kollmannsberger}, \citenamefont {Kunz}, \citenamefont
  {Biberacher}, \citenamefont {Kampert}, \citenamefont {Fujiwara},
  \citenamefont {Erb}, \citenamefont {Wosnitza}, \citenamefont {Gross},\ and\
  \citenamefont {Kartsovnik}}]{zeemannpj}%
  \BibitemOpen
  \bibfield  {author} {\bibinfo {author} {\bibfnamefont {R.}~\bibnamefont
  {Ramazashvili}}, \bibinfo {author} {\bibfnamefont {P.~D.}\ \bibnamefont
  {Grigoriev}}, \bibinfo {author} {\bibfnamefont {T.}~\bibnamefont {Helm}},
  \bibinfo {author} {\bibfnamefont {F.}~\bibnamefont {Kollmannsberger}},
  \bibinfo {author} {\bibfnamefont {M.}~\bibnamefont {Kunz}}, \bibinfo {author}
  {\bibfnamefont {W.}~\bibnamefont {Biberacher}}, \bibinfo {author}
  {\bibfnamefont {E.}~\bibnamefont {Kampert}}, \bibinfo {author} {\bibfnamefont
  {H.}~\bibnamefont {Fujiwara}}, \bibinfo {author} {\bibfnamefont
  {A.}~\bibnamefont {Erb}}, \bibinfo {author} {\bibfnamefont {J.}~\bibnamefont
  {Wosnitza}}, \bibinfo {author} {\bibfnamefont {R.}~\bibnamefont {Gross}}, \
  and\ \bibinfo {author} {\bibfnamefont {M.~V.}\ \bibnamefont {Kartsovnik}},\
  }\href@noop {} {\bibfield  {journal} {\bibinfo  {journal} {npj Quantum
  Mater.}\ }\textbf {\bibinfo {volume} {6}},\ \bibinfo {pages} {11} (\bibinfo
  {year} {2021})}\BibitemShut {NoStop}%
\bibitem [{\citenamefont {Wang}\ and\ \citenamefont
  {Qian}(2020)}]{topologafm1}%
  \BibitemOpen
  \bibfield  {author} {\bibinfo {author} {\bibfnamefont {H.}~\bibnamefont
  {Wang}}\ and\ \bibinfo {author} {\bibfnamefont {X.}~\bibnamefont {Qian}},\
  }\href@noop {} {\bibfield  {journal} {\bibinfo  {journal} {npj Comput.
  Mater.}\ }\textbf {\bibinfo {volume} {6}},\ \bibinfo {pages} {199} (\bibinfo
  {year} {2020})}\BibitemShut {NoStop}%
\bibitem [{\citenamefont {Sivadas}\ \emph {et~al.}(2016)\citenamefont
  {Sivadas}, \citenamefont {Okamoto},\ and\ \citenamefont {Xiao}}]{moke}%
  \BibitemOpen
  \bibfield  {author} {\bibinfo {author} {\bibfnamefont {N.}~\bibnamefont
  {Sivadas}}, \bibinfo {author} {\bibfnamefont {S.}~\bibnamefont {Okamoto}}, \
  and\ \bibinfo {author} {\bibfnamefont {D.}~\bibnamefont {Xiao}},\ }\href@noop
  {} {\bibfield  {journal} {\bibinfo  {journal} {Phys. Rev. Lett.}\ }\textbf
  {\bibinfo {volume} {117}},\ \bibinfo {pages} {267203} (\bibinfo {year}
  {2016})}\BibitemShut {NoStop}%
\bibitem [{\citenamefont {Fiebig}(2005)}]{mecoupling}%
  \BibitemOpen
  \bibfield  {author} {\bibinfo {author} {\bibfnamefont {M.}~\bibnamefont
  {Fiebig}},\ }\href@noop {} {\bibfield  {journal} {\bibinfo  {journal} {J.
  Phys. D: Appl. Phys.}\ }\textbf {\bibinfo {volume} {38}},\ \bibinfo {pages}
  {R123} (\bibinfo {year} {2005})}\BibitemShut {NoStop}%
\bibitem [{Note1()}]{Note1}%
  \BibitemOpen
  \bibinfo {note} {Note, however, that the coupling $\lambda _{\alpha ,\beta
  }P_\alpha \sigma _\beta $ \protect \textit {does not} exist in materials with
  time-reversal symmetry, as will be shown below.}\BibitemShut {Stop}%
\bibitem [{mpo()}]{mpoint}%
  \BibitemOpen
  \href@noop {} {\emph {\bibinfo {title} {Magnetic Point Group Tables}}}\
  (\bibinfo  {publisher}
  {https://www.cryst.ehu.es/cryst/mpoint.html})\BibitemShut {NoStop}%
\bibitem [{\citenamefont {Hergert}\ and\ \citenamefont
  {Geilhufe}(2018)}]{gtpack}%
  \BibitemOpen
  \bibfield  {author} {\bibinfo {author} {\bibfnamefont {W.}~\bibnamefont
  {Hergert}}\ and\ \bibinfo {author} {\bibfnamefont {M.}~\bibnamefont
  {Geilhufe}},\ }\href@noop {} {\emph {\bibinfo {title} {Group Theory in Solid
  State Physics and Photonics: Problem Solving with Mathematica}}}\ (\bibinfo
  {publisher} {Wiley-VCH},\ \bibinfo {year} {2018})\BibitemShut {NoStop}%
\bibitem [{\citenamefont {Koster}\ \emph {et~al.}(1963)\citenamefont {Koster},
  \citenamefont {Dimmock}, \citenamefont {Wheeler},\ and\ \citenamefont
  {Statz}}]{koster}%
  \BibitemOpen
  \bibfield  {author} {\bibinfo {author} {\bibfnamefont {G.~F.}\ \bibnamefont
  {Koster}}, \bibinfo {author} {\bibfnamefont {J.~D.}\ \bibnamefont {Dimmock}},
  \bibinfo {author} {\bibfnamefont {R.~G.}\ \bibnamefont {Wheeler}}, \ and\
  \bibinfo {author} {\bibfnamefont {H.}~\bibnamefont {Statz}},\ }\href@noop {}
  {\emph {\bibinfo {title} {Properties of the Thirty-Two Point Group}}}\
  (\bibinfo  {publisher} {M.I.T. Press},\ \bibinfo {year} {1963})\BibitemShut
  {NoStop}%
\bibitem [{\citenamefont {Altmann}\ and\ \citenamefont
  {Herzig}(2011)}]{character}%
  \BibitemOpen
  \bibfield  {author} {\bibinfo {author} {\bibfnamefont {S.~L.}\ \bibnamefont
  {Altmann}}\ and\ \bibinfo {author} {\bibfnamefont {P.}~\bibnamefont
  {Herzig}},\ }\href@noop {} {\emph {\bibinfo {title} {Point-Group Theory
  Tables (Second Edition)}}}\ (\bibinfo  {publisher} {Wien},\ \bibinfo {year}
  {2011})\BibitemShut {NoStop}%
\bibitem [{Note2()}]{Note2}%
  \BibitemOpen
  \bibinfo {note} {See Supplementary Material which includes symmetry analysis,
  methods, and some numerical results (e.g., band structures and
  orbital-projected spin magnetizations regarding SrFe$_2$S$_2$O and/or
  Fe$_2$TeO$_6$).}\BibitemShut {Stop}%
\bibitem [{poi()}]{point}%
  \BibitemOpen
  \href@noop {} {\emph {\bibinfo {title} {Point Group Tables}}}\ (\bibinfo
  {publisher} {https://www.cryst.ehu.es/rep/point.html})\BibitemShut {NoStop}%
\bibitem [{\citenamefont {Kresse}\ and\ \citenamefont
  {Furthm\"uller}(1996)}]{vasp1}%
  \BibitemOpen
  \bibfield  {author} {\bibinfo {author} {\bibfnamefont {G.}~\bibnamefont
  {Kresse}}\ and\ \bibinfo {author} {\bibfnamefont {J.}~\bibnamefont
  {Furthm\"uller}},\ }\href@noop {} {\bibfield  {journal} {\bibinfo  {journal}
  {Phys. Rev. B}\ }\textbf {\bibinfo {volume} {54}},\ \bibinfo {pages} {11169}
  (\bibinfo {year} {1996})}\BibitemShut {NoStop}%
\bibitem [{\citenamefont {Kresse}\ and\ \citenamefont {Joubert}(1999)}]{vasp2}%
  \BibitemOpen
  \bibfield  {author} {\bibinfo {author} {\bibfnamefont {G.}~\bibnamefont
  {Kresse}}\ and\ \bibinfo {author} {\bibfnamefont {D.}~\bibnamefont
  {Joubert}},\ }\href@noop {} {\bibfield  {journal} {\bibinfo  {journal} {Phys.
  Rev. B}\ }\textbf {\bibinfo {volume} {59}},\ \bibinfo {pages} {1758}
  (\bibinfo {year} {1999})}\BibitemShut {NoStop}%
\bibitem [{\citenamefont {Bl\"ochl}(1994)}]{paw}%
  \BibitemOpen
  \bibfield  {author} {\bibinfo {author} {\bibfnamefont {P.~E.}\ \bibnamefont
  {Bl\"ochl}},\ }\href@noop {} {\bibfield  {journal} {\bibinfo  {journal}
  {Phys. Rev. B}\ }\textbf {\bibinfo {volume} {50}},\ \bibinfo {pages} {17953}
  (\bibinfo {year} {1994})}\BibitemShut {NoStop}%
\bibitem [{\citenamefont {Ceperley}\ and\ \citenamefont {Alder}(1980)}]{ldaca}%
  \BibitemOpen
  \bibfield  {author} {\bibinfo {author} {\bibfnamefont {D.~M.}\ \bibnamefont
  {Ceperley}}\ and\ \bibinfo {author} {\bibfnamefont {B.~J.}\ \bibnamefont
  {Alder}},\ }\href@noop {} {\bibfield  {journal} {\bibinfo  {journal} {Phys.
  Rev. Lett.}\ }\textbf {\bibinfo {volume} {45}},\ \bibinfo {pages} {566}
  (\bibinfo {year} {1980})}\BibitemShut {NoStop}%
\bibitem [{\citenamefont {Dudarev}\ \emph {et~al.}(1998)\citenamefont
  {Dudarev}, \citenamefont {Botton}, \citenamefont {Savrasov}, \citenamefont
  {Humphreys},\ and\ \citenamefont {Sutton}}]{ldau}%
  \BibitemOpen
  \bibfield  {author} {\bibinfo {author} {\bibfnamefont {S.~L.}\ \bibnamefont
  {Dudarev}}, \bibinfo {author} {\bibfnamefont {G.~A.}\ \bibnamefont {Botton}},
  \bibinfo {author} {\bibfnamefont {S.~Y.}\ \bibnamefont {Savrasov}}, \bibinfo
  {author} {\bibfnamefont {C.~J.}\ \bibnamefont {Humphreys}}, \ and\ \bibinfo
  {author} {\bibfnamefont {A.~P.}\ \bibnamefont {Sutton}},\ }\href {\doibase
  10.1103/PhysRevB.57.1505} {\bibfield  {journal} {\bibinfo  {journal} {Phys.
  Rev. B}\ }\textbf {\bibinfo {volume} {57}},\ \bibinfo {pages} {1505}
  (\bibinfo {year} {1998})}\BibitemShut {NoStop}%
\bibitem [{\citenamefont {Inc.}()}]{mma}%
  \BibitemOpen
  \bibfield  {author} {\bibinfo {author} {\bibfnamefont {W.~R.}\ \bibnamefont
  {Inc.}},\ }\href {https://www.wolfram.com/mathematica} {\enquote {\bibinfo
  {title} {Mathematica, {V}ersion 12.0},}\ }\bibinfo {note} {Champaign, IL,
  2019}\BibitemShut {NoStop}%
\bibitem [{mat()}]{materialsproject}%
  \BibitemOpen
  \href@noop {} {\emph {\bibinfo {title} {Materials Project}}}\ (\bibinfo
  {publisher} {https://materialsproject.org/})\BibitemShut {NoStop}%
\bibitem [{\citenamefont {Jain}\ \emph {et~al.}(2013)\citenamefont {Jain},
  \citenamefont {Ong}, \citenamefont {Hautier}, \citenamefont {Chen},
  \citenamefont {Richards}, \citenamefont {Dacek}, \citenamefont {Cholia},
  \citenamefont {Gunter}, \citenamefont {Skinner}, \citenamefont {Ceder},\ and\
  \citenamefont {Persson}}]{materialsproject2}%
  \BibitemOpen
  \bibfield  {author} {\bibinfo {author} {\bibfnamefont {A.}~\bibnamefont
  {Jain}}, \bibinfo {author} {\bibfnamefont {S.~P.}\ \bibnamefont {Ong}},
  \bibinfo {author} {\bibfnamefont {G.}~\bibnamefont {Hautier}}, \bibinfo
  {author} {\bibfnamefont {W.}~\bibnamefont {Chen}}, \bibinfo {author}
  {\bibfnamefont {W.~D.}\ \bibnamefont {Richards}}, \bibinfo {author}
  {\bibfnamefont {S.}~\bibnamefont {Dacek}}, \bibinfo {author} {\bibfnamefont
  {S.}~\bibnamefont {Cholia}}, \bibinfo {author} {\bibfnamefont
  {D.}~\bibnamefont {Gunter}}, \bibinfo {author} {\bibfnamefont
  {D.}~\bibnamefont {Skinner}}, \bibinfo {author} {\bibfnamefont
  {G.}~\bibnamefont {Ceder}}, \ and\ \bibinfo {author} {\bibfnamefont {K.~A.}\
  \bibnamefont {Persson}},\ }\href {\doibase 10.1063/1.4812323} {\bibfield
  {journal} {\bibinfo  {journal} {APL Mater.}\ }\textbf {\bibinfo {volume}
  {1}},\ \bibinfo {pages} {011002} (\bibinfo {year} {2013})}\BibitemShut
  {NoStop}%
\bibitem [{\citenamefont {Momma}\ and\ \citenamefont {Izumi}(2011)}]{vesta}%
  \BibitemOpen
  \bibfield  {author} {\bibinfo {author} {\bibfnamefont {K.}~\bibnamefont
  {Momma}}\ and\ \bibinfo {author} {\bibfnamefont {F.}~\bibnamefont {Izumi}},\
  }\href {\doibase 10.1107/S0021889811038970} {\bibfield  {journal} {\bibinfo
  {journal} {J. Appl. Crystal.}\ }\textbf {\bibinfo {volume} {44}},\ \bibinfo
  {pages} {1272} (\bibinfo {year} {2011})}\BibitemShut {NoStop}%
\bibitem [{\citenamefont {Wang}\ \emph {et~al.}(2021)\citenamefont {Wang},
  \citenamefont {Xu}, \citenamefont {Liu}, \citenamefont {Tang},\ and\
  \citenamefont {Geng}}]{vaspkit}%
  \BibitemOpen
  \bibfield  {author} {\bibinfo {author} {\bibfnamefont {V.}~\bibnamefont
  {Wang}}, \bibinfo {author} {\bibfnamefont {N.}~\bibnamefont {Xu}}, \bibinfo
  {author} {\bibfnamefont {J.-C.}\ \bibnamefont {Liu}}, \bibinfo {author}
  {\bibfnamefont {G.}~\bibnamefont {Tang}}, \ and\ \bibinfo {author}
  {\bibfnamefont {W.-T.}\ \bibnamefont {Geng}},\ }\href {\doibase
  https://doi.org/10.1016/j.cpc.2021.108033} {\bibfield  {journal} {\bibinfo
  {journal} {Comput. Phys. Commun.}\ }\textbf {\bibinfo {volume} {267}},\
  \bibinfo {pages} {108033} (\bibinfo {year} {2021})}\BibitemShut {NoStop}%
\bibitem [{vas()}]{vaspkit2}%
  \BibitemOpen
  \href@noop {} {\emph {\bibinfo {title} {VASPKIT}}}\ (\bibinfo  {publisher}
  {https://vaspkit.com})\BibitemShut {NoStop}%
\bibitem [{\citenamefont {Hunter}(2007)}]{matplotlib}%
  \BibitemOpen
  \bibfield  {author} {\bibinfo {author} {\bibfnamefont {J.~D.}\ \bibnamefont
  {Hunter}},\ }\href {\doibase 10.1109/MCSE.2007.55} {\bibfield  {journal}
  {\bibinfo  {journal} {Comput. Sci. Eng.}\ }\textbf {\bibinfo {volume} {9}},\
  \bibinfo {pages} {90} (\bibinfo {year} {2007})}\BibitemShut {NoStop}%
\bibitem [{see()}]{seekpath}%
  \BibitemOpen
  \href@noop {} {\emph {\bibinfo {title} {SeeK-path}}}\ (\bibinfo  {publisher}
  {https://www.materialscloud.org/work/tools/seekpath})\BibitemShut {NoStop}%
\bibitem [{\citenamefont {Hinuma}\ \emph {et~al.}(2017)\citenamefont {Hinuma},
  \citenamefont {Pizzi}, \citenamefont {Kumagai}, \citenamefont {Oba},\ and\
  \citenamefont {Tanaka}}]{seekpath2}%
  \BibitemOpen
  \bibfield  {author} {\bibinfo {author} {\bibfnamefont {Y.}~\bibnamefont
  {Hinuma}}, \bibinfo {author} {\bibfnamefont {G.}~\bibnamefont {Pizzi}},
  \bibinfo {author} {\bibfnamefont {Y.}~\bibnamefont {Kumagai}}, \bibinfo
  {author} {\bibfnamefont {F.}~\bibnamefont {Oba}}, \ and\ \bibinfo {author}
  {\bibfnamefont {I.}~\bibnamefont {Tanaka}},\ }\href {\doibase
  10.1016/j.commatsci.2016.10.015} {\bibfield  {journal} {\bibinfo  {journal}
  {Comp. Mater. Sci.}\ }\textbf {\bibinfo {volume} {128}},\ \bibinfo {pages}
  {140} (\bibinfo {year} {2017})}\BibitemShut {NoStop}%
\bibitem [{\citenamefont {Togo}\ and\ \citenamefont
  {Tanaka}(2018)}]{seekpath3}%
  \BibitemOpen
  \bibfield  {author} {\bibinfo {author} {\bibfnamefont {A.}~\bibnamefont
  {Togo}}\ and\ \bibinfo {author} {\bibfnamefont {I.}~\bibnamefont {Tanaka}},\
  }\href@noop {} {\enquote {\bibinfo {title} {$\texttt{Spglib}$: a software
  library for crystal symmetry search},}\ } (\bibinfo {year} {2018}),\ \Eprint
  {http://arxiv.org/abs/arXiv:1808.01590} {arXiv:1808.01590} \BibitemShut
  {NoStop}%
\bibitem [{\citenamefont {Stokes}\ and\ \citenamefont
  {Hatch}(2005)}]{findsym1}%
  \BibitemOpen
  \bibfield  {author} {\bibinfo {author} {\bibfnamefont {H.~T.}\ \bibnamefont
  {Stokes}}\ and\ \bibinfo {author} {\bibfnamefont {D.~M.}\ \bibnamefont
  {Hatch}},\ }\href@noop {} {\bibfield  {journal} {\bibinfo  {journal} {J.
  Appl. Cryst.}\ }\textbf {\bibinfo {volume} {38}},\ \bibinfo {pages} {237}
  (\bibinfo {year} {2005})}\BibitemShut {NoStop}%
\bibitem [{fin()}]{findsym2}%
  \BibitemOpen
  \href@noop {} {\emph {\bibinfo {title} {FINDSYM}}}\ (\bibinfo  {publisher}
  {https://stokes.byu.edu/iso/findsym.php})\BibitemShut {NoStop}%
\bibitem [{\citenamefont {Aroyo}\ \emph
  {et~al.}(2006{\natexlab{a}})\citenamefont {Aroyo}, \citenamefont {Kirov},
  \citenamefont {Capillas}, \citenamefont {Perez-Mato},\ and\ \citenamefont
  {Wondratschek}}]{aroyo2006}%
  \BibitemOpen
  \bibfield  {author} {\bibinfo {author} {\bibfnamefont {M.~I.}\ \bibnamefont
  {Aroyo}}, \bibinfo {author} {\bibfnamefont {A.}~\bibnamefont {Kirov}},
  \bibinfo {author} {\bibfnamefont {C.}~\bibnamefont {Capillas}}, \bibinfo
  {author} {\bibfnamefont {J.~M.}\ \bibnamefont {Perez-Mato}}, \ and\ \bibinfo
  {author} {\bibfnamefont {H.}~\bibnamefont {Wondratschek}},\ }\href {\doibase
  10.1107/s0108767305040286} {\bibfield  {journal} {\bibinfo  {journal} {Acta
  Cryst. A}\ }\textbf {\bibinfo {volume} {62}},\ \bibinfo {pages} {115}
  (\bibinfo {year} {2006}{\natexlab{a}})}\BibitemShut {NoStop}%
\bibitem [{\citenamefont {Aroyo}\ \emph
  {et~al.}(2006{\natexlab{b}})\citenamefont {Aroyo}, \citenamefont
  {Perez-Mato}, \citenamefont {Capillas}, \citenamefont {Kroumova},
  \citenamefont {Ivantchev}, \citenamefont {Madariaga}, \citenamefont {Kirov},\
  and\ \citenamefont {Wondratschek}}]{aroyo20062}%
  \BibitemOpen
  \bibfield  {author} {\bibinfo {author} {\bibfnamefont {M.~I.}\ \bibnamefont
  {Aroyo}}, \bibinfo {author} {\bibfnamefont {J.~M.}\ \bibnamefont
  {Perez-Mato}}, \bibinfo {author} {\bibfnamefont {C.}~\bibnamefont
  {Capillas}}, \bibinfo {author} {\bibfnamefont {E.}~\bibnamefont {Kroumova}},
  \bibinfo {author} {\bibfnamefont {S.}~\bibnamefont {Ivantchev}}, \bibinfo
  {author} {\bibfnamefont {G.}~\bibnamefont {Madariaga}}, \bibinfo {author}
  {\bibfnamefont {A.}~\bibnamefont {Kirov}}, \ and\ \bibinfo {author}
  {\bibfnamefont {H.}~\bibnamefont {Wondratschek}},\ }\href {\doibase
  10.1524/zkri.2006.221.1.15} {\bibfield  {journal} {\bibinfo  {journal} {Z.
  Kristallogr. Cryst. Mater.}\ }\textbf {\bibinfo {volume} {221}},\ \bibinfo
  {pages} {15} (\bibinfo {year} {2006}{\natexlab{b}})}\BibitemShut {NoStop}%
\bibitem [{mag({\natexlab{a}})}]{magndata}%
  \BibitemOpen
  \href@noop {} {\emph {\bibinfo {title} {MAGNDATA}}}\ (\bibinfo  {publisher}
  {http://webbdcrista1.ehu.es/magndata})\BibitemShut {NoStop}%
\bibitem [{\citenamefont {Gallego}\ \emph
  {et~al.}(2016{\natexlab{a}})\citenamefont {Gallego}, \citenamefont
  {Perez-Mato}, \citenamefont {Elcoro}, \citenamefont {Tasci}, \citenamefont
  {Hanson}, \citenamefont {Momma}, \citenamefont {Aroyo},\ and\ \citenamefont
  {Madariaga}}]{magndata2}%
  \BibitemOpen
  \bibfield  {author} {\bibinfo {author} {\bibfnamefont {S.~V.}\ \bibnamefont
  {Gallego}}, \bibinfo {author} {\bibfnamefont {J.~M.}\ \bibnamefont
  {Perez-Mato}}, \bibinfo {author} {\bibfnamefont {L.}~\bibnamefont {Elcoro}},
  \bibinfo {author} {\bibfnamefont {E.~S.}\ \bibnamefont {Tasci}}, \bibinfo
  {author} {\bibfnamefont {R.~M.}\ \bibnamefont {Hanson}}, \bibinfo {author}
  {\bibfnamefont {K.}~\bibnamefont {Momma}}, \bibinfo {author} {\bibfnamefont
  {M.~I.}\ \bibnamefont {Aroyo}}, \ and\ \bibinfo {author} {\bibfnamefont
  {G.}~\bibnamefont {Madariaga}},\ }\href@noop {} {\bibfield  {journal}
  {\bibinfo  {journal} {J. Appl. Crystallogr.}\ }\textbf {\bibinfo {volume}
  {49}},\ \bibinfo {pages} {1750} (\bibinfo {year}
  {2016}{\natexlab{a}})}\BibitemShut {NoStop}%
\bibitem [{\citenamefont {Gallego}\ \emph
  {et~al.}(2016{\natexlab{b}})\citenamefont {Gallego}, \citenamefont
  {Perez-Mato}, \citenamefont {Elcoro}, \citenamefont {Tasci}, \citenamefont
  {Hanson}, \citenamefont {Aroyo},\ and\ \citenamefont
  {Madariaga}}]{magndata3}%
  \BibitemOpen
  \bibfield  {author} {\bibinfo {author} {\bibfnamefont {S.~V.}\ \bibnamefont
  {Gallego}}, \bibinfo {author} {\bibfnamefont {J.~M.}\ \bibnamefont
  {Perez-Mato}}, \bibinfo {author} {\bibfnamefont {L.}~\bibnamefont {Elcoro}},
  \bibinfo {author} {\bibfnamefont {E.~S.}\ \bibnamefont {Tasci}}, \bibinfo
  {author} {\bibfnamefont {R.~M.}\ \bibnamefont {Hanson}}, \bibinfo {author}
  {\bibfnamefont {M.~I.}\ \bibnamefont {Aroyo}}, \ and\ \bibinfo {author}
  {\bibfnamefont {G.}~\bibnamefont {Madariaga}},\ }\href@noop {} {\bibfield
  {journal} {\bibinfo  {journal} {J. Appl. Crystallogr.}\ }\textbf {\bibinfo
  {volume} {49}},\ \bibinfo {pages} {1941} (\bibinfo {year}
  {2016}{\natexlab{b}})}\BibitemShut {NoStop}%
\bibitem [{\citenamefont {Herath}\ \emph {et~al.}(2020)\citenamefont {Herath},
  \citenamefont {Tavadze}, \citenamefont {He}, \citenamefont {Bousquet},
  \citenamefont {Singh}, \citenamefont {Munoz},\ and\ \citenamefont
  {Romero}}]{pyprocar}%
  \BibitemOpen
  \bibfield  {author} {\bibinfo {author} {\bibfnamefont {U.}~\bibnamefont
  {Herath}}, \bibinfo {author} {\bibfnamefont {P.}~\bibnamefont {Tavadze}},
  \bibinfo {author} {\bibfnamefont {X.}~\bibnamefont {He}}, \bibinfo {author}
  {\bibfnamefont {E.}~\bibnamefont {Bousquet}}, \bibinfo {author}
  {\bibfnamefont {S.}~\bibnamefont {Singh}}, \bibinfo {author} {\bibfnamefont
  {F.}~\bibnamefont {Munoz}}, \ and\ \bibinfo {author} {\bibfnamefont {A.~H.}\
  \bibnamefont {Romero}},\ }\href@noop {} {\bibfield  {journal} {\bibinfo
  {journal} {Comput. Phys. Commun.}\ }\textbf {\bibinfo {volume} {251}},\
  \bibinfo {pages} {107080} (\bibinfo {year} {2020})}\BibitemShut {NoStop}%
\bibitem [{pyp()}]{pyprocar2}%
  \BibitemOpen
  \href@noop {} {\emph {\bibinfo {title} {PYPROCAR}}}\ (\bibinfo  {publisher}
  {https://romerogroup.github.io/pyprocar/index.html})\BibitemShut {NoStop}%
\bibitem [{\citenamefont {Manchon}\ \emph {et~al.}(2015)\citenamefont
  {Manchon}, \citenamefont {Koo}, \citenamefont {Nitta}, \citenamefont
  {Frolov},\ and\ \citenamefont {Duine}}]{rashbasoc}%
  \BibitemOpen
  \bibfield  {author} {\bibinfo {author} {\bibfnamefont {A.}~\bibnamefont
  {Manchon}}, \bibinfo {author} {\bibfnamefont {H.~C.}\ \bibnamefont {Koo}},
  \bibinfo {author} {\bibfnamefont {J.}~\bibnamefont {Nitta}}, \bibinfo
  {author} {\bibfnamefont {S.~M.}\ \bibnamefont {Frolov}}, \ and\ \bibinfo
  {author} {\bibfnamefont {R.~A.}\ \bibnamefont {Duine}},\ }\href@noop {}
  {\bibfield  {journal} {\bibinfo  {journal} {Nat. Mater.}\ }\textbf {\bibinfo
  {volume} {14}},\ \bibinfo {pages} {871} (\bibinfo {year} {2015})}\BibitemShut
  {NoStop}%
\bibitem [{mag({\natexlab{b}})}]{magnetoelectric}%
  \BibitemOpen
  \href@noop {} {\emph {\bibinfo {title} {MTENSOR}}}\ (\bibinfo  {publisher}
  {https://www.cryst.ehu.es/cgi-bin/cryst/programs/mtensor.pl})\BibitemShut
  {NoStop}%
\bibitem [{\citenamefont {{\v{S}}mejkal}\ \emph {et~al.}(2018)\citenamefont
  {{\v{S}}mejkal}, \citenamefont {Mokrousov}, \citenamefont {Yan},\ and\
  \citenamefont {MacDonald}}]{topologafm2}%
  \BibitemOpen
  \bibfield  {author} {\bibinfo {author} {\bibfnamefont {L.}~\bibnamefont
  {{\v{S}}mejkal}}, \bibinfo {author} {\bibfnamefont {Y.}~\bibnamefont
  {Mokrousov}}, \bibinfo {author} {\bibfnamefont {B.}~\bibnamefont {Yan}}, \
  and\ \bibinfo {author} {\bibfnamefont {A.~H.}\ \bibnamefont {MacDonald}},\
  }\href {\doibase 10.1038/s41567-018-0064-5} {\bibfield  {journal} {\bibinfo
  {journal} {Nat. Phys.}\ }\textbf {\bibinfo {volume} {14}},\ \bibinfo {pages}
  {242} (\bibinfo {year} {2018})}\BibitemShut {NoStop}%
\bibitem [{\citenamefont {Voon}\ \emph {et~al.}(2009)\citenamefont {Voon},
  \citenamefont {C.},\ and\ \citenamefont {Willatzen}}]{kpmethod2009}%
  \BibitemOpen
  \bibfield  {author} {\bibinfo {author} {\bibfnamefont {L.~Y.}\ \bibnamefont
  {Voon}}, \bibinfo {author} {\bibfnamefont {L.}~\bibnamefont {C.}}, \ and\
  \bibinfo {author} {\bibfnamefont {M.}~\bibnamefont {Willatzen}},\ }\href@noop
  {} {\emph {\bibinfo {title} {The $k\cdot p$ Method: Electronic Properties of
  Semiconductors}}}\ (\bibinfo  {publisher} {Springer, Berlin},\ \bibinfo
  {year} {2009})\BibitemShut {NoStop}%
\bibitem [{\citenamefont {Dresselhaus}\ \emph {et~al.}(2008)\citenamefont
  {Dresselhaus}, \citenamefont {Dresselhaus},\ and\ \citenamefont
  {Jario}}]{grouptheorydresselhaus}%
  \BibitemOpen
  \bibfield  {author} {\bibinfo {author} {\bibfnamefont {M.}~\bibnamefont
  {Dresselhaus}}, \bibinfo {author} {\bibfnamefont {G.}~\bibnamefont
  {Dresselhaus}}, \ and\ \bibinfo {author} {\bibfnamefont {A.}~\bibnamefont
  {Jario}},\ }\href {\doibase 10.1007/978-3-540-32899-5} {\emph {\bibinfo
  {title} {Group Theory -- Application to the Physics of Condensed Matter}}}\
  (\bibinfo  {publisher} {Springer-Verlag Berlin Heidelberg},\ \bibinfo {year}
  {2008})\BibitemShut {NoStop}%
\bibitem [{Note3()}]{Note3}%
  \BibitemOpen
  \bibinfo {note} {Here, the second equality holds because that $\protect
  \mathcal {E}_\alpha $ polarizes centrosymmetric materials by creating
  $P_\alpha $.}\BibitemShut {Stop}%
\bibitem [{\citenamefont {Guo}\ \emph {et~al.}(2017)\citenamefont {Guo},
  \citenamefont {Fern\'{a}ndez-D\'{i}az}, \citenamefont {Komarek},
  \citenamefont {Huh}, \citenamefont {Adler},\ and\ \citenamefont
  {Valldor}}]{magstr}%
  \BibitemOpen
  \bibfield  {author} {\bibinfo {author} {\bibfnamefont {H.}~\bibnamefont
  {Guo}}, \bibinfo {author} {\bibfnamefont {M.-T.}\ \bibnamefont
  {Fern\'{a}ndez-D\'{i}az}}, \bibinfo {author} {\bibfnamefont {A.~C.}\
  \bibnamefont {Komarek}}, \bibinfo {author} {\bibfnamefont {S.}~\bibnamefont
  {Huh}}, \bibinfo {author} {\bibfnamefont {P.}~\bibnamefont {Adler}}, \ and\
  \bibinfo {author} {\bibfnamefont {M.}~\bibnamefont {Valldor}},\ }\href@noop
  {} {\bibfield  {journal} {\bibinfo  {journal} {Eur. J. Inorg. Chem.}\
  }\textbf {\bibinfo {volume} {2017}},\ \bibinfo {pages} {3829} (\bibinfo
  {year} {2017})}\BibitemShut {NoStop}%
\bibitem [{\citenamefont {Kunnmann}\ \emph {et~al.}(1968)\citenamefont
  {Kunnmann}, \citenamefont {Placa}, \citenamefont {Corliss}, \citenamefont
  {Hastings},\ and\ \citenamefont {Banks}}]{fe2teo6}%
  \BibitemOpen
  \bibfield  {author} {\bibinfo {author} {\bibfnamefont {W.}~\bibnamefont
  {Kunnmann}}, \bibinfo {author} {\bibfnamefont {S.~L.}\ \bibnamefont {Placa}},
  \bibinfo {author} {\bibfnamefont {L.}~\bibnamefont {Corliss}}, \bibinfo
  {author} {\bibfnamefont {J.}~\bibnamefont {Hastings}}, \ and\ \bibinfo
  {author} {\bibfnamefont {E.}~\bibnamefont {Banks}},\ }\href@noop {}
  {\bibfield  {journal} {\bibinfo  {journal} {J. Phys. Chem. Solids}\ }\textbf
  {\bibinfo {volume} {29}},\ \bibinfo {pages} {1359} (\bibinfo {year}
  {1968})}\BibitemShut {NoStop}%
\bibitem [{\citenamefont {Buksphan}\ \emph {et~al.}(1972)\citenamefont
  {Buksphan}, \citenamefont {Fischer},\ and\ \citenamefont
  {Hornreich}}]{fe2teo62}%
  \BibitemOpen
  \bibfield  {author} {\bibinfo {author} {\bibfnamefont {S.}~\bibnamefont
  {Buksphan}}, \bibinfo {author} {\bibfnamefont {E.}~\bibnamefont {Fischer}}, \
  and\ \bibinfo {author} {\bibfnamefont {R.}~\bibnamefont {Hornreich}},\
  }\href@noop {} {\bibfield  {journal} {\bibinfo  {journal} {Solid State
  Commun.}\ }\textbf {\bibinfo {volume} {10}},\ \bibinfo {pages} {657}
  (\bibinfo {year} {1972})}\BibitemShut {NoStop}%
\bibitem [{Note4()}]{Note4}%
  \BibitemOpen
  \bibinfo {note} {We also numerically found that polarizing Fe$_2$TeO$_6$ by
  $\protect \mathcal {E}_x=6$ MV/cm causes a tiny Zeeman spin splitting of
  $\sim $3 meV at the CBM.}\BibitemShut {Stop}%
\bibitem [{Note5()}]{Note5}%
  \BibitemOpen
  \bibinfo {note} {The conventional Zeeman spin splitting created by magnetic
  field $B$ is given by $\lvert g^\prime \rvert \mu _B B$, where $g^\prime $ is
  the effective Land\'e $g$-factor~\cite {spintronic,spintronic2}. For free
  electrons or electrons in non-magnetic semiconductors with large enough band
  gap, the $g^\prime $ is nearly 2.0~\cite {spintronic,spintronic2}. The Zeeman
  spin splitting of $\sim $55 meV can be driven by magnetic field of $\sim $475
  Tesla. In II-V wurtzite semiconductors ZnS and CdSe, the effective $\lvert
  g^\prime \rvert $ are 0.6 and 2.3, respectively~\cite {spintronic2}.
  Generating Zeeman splitting of $\sim $55 meV in ZnS and CdSe semiconductors
  thus requires a magnetic field of $\sim $413 and $\sim $1583 Tesla,
  respectively. In III-V semiconductor InSb, the effective $\lvert g^\prime
  \rvert $ can be as large as 51.3~\cite {spintronic2}. In such a case,
  magnetic field of $\sim $18 Tesla creates Zeeman spin splitting of $\sim $55
  meV.}\BibitemShut {Stop}%
\bibitem [{\citenamefont {Fu}\ and\ \citenamefont {Bellaiche}(2003)}]{efield}%
  \BibitemOpen
  \bibfield  {author} {\bibinfo {author} {\bibfnamefont {H.}~\bibnamefont
  {Fu}}\ and\ \bibinfo {author} {\bibfnamefont {L.}~\bibnamefont {Bellaiche}},\
  }\href@noop {} {\bibfield  {journal} {\bibinfo  {journal} {Phys. Rev. Lett.}\
  }\textbf {\bibinfo {volume} {91}},\ \bibinfo {pages} {057601} (\bibinfo
  {year} {2003})}\BibitemShut {NoStop}%
\bibitem [{Note6()}]{Note6}%
  \BibitemOpen
  \bibinfo {note} {Given the spinor quantum state $\mathinner {|{\psi }\rangle
  }$, the expectation value of spin magnetization $S_\alpha $ is defined by
  $\protect \frac {1}{2}\mathinner {\langle {\psi }|}\sigma _\alpha \mathinner
  {|{\psi }\rangle }$ ($\alpha =x,y,z$). See, e.g., Ref.~\cite
  {rashbadresselhaus2}.}\BibitemShut {Stop}%
\bibitem [{Note7()}]{Note7}%
  \BibitemOpen
  \bibinfo {note} {Neglecting the spin-orbit interaction does not qualitatively
  change our predictions for SrFe$_2$S$_2$O and Fe$_2$TeO$_6$ (see Fig.~S6 of
  the SM).}\BibitemShut {Stop}%
\bibitem [{Note8()}]{Note8}%
  \BibitemOpen
  \bibinfo {note} {As for SrFe$_2$S$_2$O polarized by $\protect \mathcal
  {E}_x=\pm 6$ MV/cm, the induced spin splittings are too tiny to
  detect.}\BibitemShut {Stop}%
\bibitem [{\citenamefont {Zhang}\ \emph {et~al.}(2018)\citenamefont {Zhang},
  \citenamefont {Li}, \citenamefont {Cao}, \citenamefont {Xiao}, \citenamefont
  {Guo},\ and\ \citenamefont {Guo}}]{spintronic3}%
  \BibitemOpen
  \bibfield  {author} {\bibinfo {author} {\bibfnamefont {X.}~\bibnamefont
  {Zhang}}, \bibinfo {author} {\bibfnamefont {H.-O.}\ \bibnamefont {Li}},
  \bibinfo {author} {\bibfnamefont {G.}~\bibnamefont {Cao}}, \bibinfo {author}
  {\bibfnamefont {M.}~\bibnamefont {Xiao}}, \bibinfo {author} {\bibfnamefont
  {G.-C.}\ \bibnamefont {Guo}}, \ and\ \bibinfo {author} {\bibfnamefont
  {G.-P.}\ \bibnamefont {Guo}},\ }\href@noop {} {\bibfield  {journal} {\bibinfo
   {journal} {Natl. Sci. Rev.}\ }\textbf {\bibinfo {volume} {6}},\ \bibinfo
  {pages} {32} (\bibinfo {year} {2018})}\BibitemShut {NoStop}%
\bibitem [{\citenamefont {Baltz}\ \emph {et~al.}(2018)\citenamefont {Baltz},
  \citenamefont {Manchon}, \citenamefont {Tsoi}, \citenamefont {Moriyama},
  \citenamefont {Ono},\ and\ \citenamefont {Tserkovnyak}}]{afm}%
  \BibitemOpen
  \bibfield  {author} {\bibinfo {author} {\bibfnamefont {V.}~\bibnamefont
  {Baltz}}, \bibinfo {author} {\bibfnamefont {A.}~\bibnamefont {Manchon}},
  \bibinfo {author} {\bibfnamefont {M.}~\bibnamefont {Tsoi}}, \bibinfo {author}
  {\bibfnamefont {T.}~\bibnamefont {Moriyama}}, \bibinfo {author}
  {\bibfnamefont {T.}~\bibnamefont {Ono}}, \ and\ \bibinfo {author}
  {\bibfnamefont {Y.}~\bibnamefont {Tserkovnyak}},\ }\href {\doibase
  10.1103/RevModPhys.90.015005} {\bibfield  {journal} {\bibinfo  {journal}
  {Rev. Mod. Phys.}\ }\textbf {\bibinfo {volume} {90}},\ \bibinfo {pages}
  {015005} (\bibinfo {year} {2018})}\BibitemShut {NoStop}%
\bibitem [{\citenamefont {Gomonay}\ \emph {et~al.}(2018)\citenamefont
  {Gomonay}, \citenamefont {Baltz}, \citenamefont {Brataas},\ and\
  \citenamefont {Tserkovnyak}}]{afmspin3}%
  \BibitemOpen
  \bibfield  {author} {\bibinfo {author} {\bibfnamefont {O.}~\bibnamefont
  {Gomonay}}, \bibinfo {author} {\bibfnamefont {V.}~\bibnamefont {Baltz}},
  \bibinfo {author} {\bibfnamefont {A.}~\bibnamefont {Brataas}}, \ and\
  \bibinfo {author} {\bibfnamefont {Y.}~\bibnamefont {Tserkovnyak}},\ }\href
  {\doibase 10.1038/s41567-018-0049-4} {\bibfield  {journal} {\bibinfo
  {journal} {Nat. Phys.}\ }\textbf {\bibinfo {volume} {14}},\ \bibinfo {pages}
  {213} (\bibinfo {year} {2018})}\BibitemShut {NoStop}%
\end{thebibliography}
%

\end{document}